\documentclass[article,3p]{elsarticle}

\usepackage{ecrc}
\usepackage{placeins}
\usepackage[T1]{fontenc}
\usepackage{subfig}
\newtheorem{definition}{Definition}

\volume{00}

\firstpage{1}

\journalname{Structural Safety}

\runauth{Max Ehre, Iason Papaioannou, Daniel Straub}

\jid{procs}

\jnltitlelogo{Structural Safety}

\CopyrightLine{2011}{Published by Elsevier Ltd.}

\usepackage{amssymb}
\usepackage[figuresright]{rotating}
\usepackage{bm}
\usepackage{mathtools}
\usepackage{cleveref}

\crefname{equation}{Eq.}{Eqs.}
\crefname{figure}{Fig.}{Figs.}
\crefname{tabular}{Tab.}{Tabs.}
\crefname{table}{Tab.}{Tabs.}
\crefname{section}{Section}{Sections}
\crefname{subsection}{Subsection}{Subsections}
\crefname{algorithm}{Alg.}{Algs.}
\crefname{chapter}{Chapter}{Chapters}
\crefname{definition}{Definition}{Definitions}
\crefname{remark}{Remark}{Remarks}

\usepackage{lineno}

\let\oldalign\align
\let\oldendalign\endalign

\renewenvironment{align}
{\linenomathNonumbers\oldalign}
{\oldendalign\endlinenomath}

\makeatletter
\renewcommand*{\@opargbegintheorem}[3]{\trivlist
	\item[\hskip \labelsep{\bfseries #1\ #2}] \textbf{(#3)}\ \itshape}
\makeatother

\begin{document}
\begin{frontmatter}



\dochead{}

\title{Variance-based reliability sensitivity with dependent inputs using failure samples}

\author[label1]{Max Ehre}
\author[label1]{Iason Papaioannou}
\author[label1]{Daniel Straub}
\address[label1]{Engineering Risk Analysis Group, TU M{\"u}nchen, Theresienstrasse 90, 80290 Munich}

\begin{abstract}
Reliability sensitivity analysis is concerned with measuring the influence of a system's uncertain input parameters on its probability of failure. Statistically dependent inputs present a challenge in both computing and interpreting these sensitivity indices; such dependencies require discerning between variable interactions produced by the probabilistic model describing the system inputs and the computational model describing the system itself.
To accomplish such a separation of effects in the context of reliability sensitivity analysis we extend on an idea originally proposed by Mara and Tarantola \cite{Mara2012} for model outputs unrelated to rare events. We compute the independent (influence via computational model) and full (influence via both computational and probabilistic model) contributions of all inputs to the variance of the indicator function of the rare event. We compute this full set of variance-based sensitivity indices of the rare event indicator using a single set of failure samples. This is possible by considering $d$ different hierarchically structured isoprobabilistic transformations of this set of failure samples from the original $d$-dimensional space of dependent inputs to standard-normal space. 
The approach facilitates computing the full set of variance-based reliability sensitivity indices with a single set of failure samples obtained as the byproduct of a single run of a sample-based rare event estimation method. That is, no additional evaluations of the computational model are required. We demonstrate the approach on a test function and two engineering problems.
\end{abstract}

\begin{keyword}
keyword \sep keyword



\end{keyword}

\end{frontmatter}
\section{Introduction}
\label{sec:intro}
When considering the safety of an engineering system under uncertainty, the key performance indicator is the probability of failure. Let the system be described by some (computational) model \(\mathcal{Y}: \bm{X} \to Y\). All uncertain parameters are gathered in the input $\bm{X}$ and $\mathcal{Y}$ is a deterministic map. 
We assume $\bm{X}: \Omega \to \mathcal{X} \subseteq \mathbb{R}^d$ and $Y: \Omega \to \mathcal{Y} \subseteq  \mathbb{R}$ to be absolutely continuous random vectors/variables mapping from the sample space $\Omega$  to the outcome spaces $\mathcal{X}$ and $\mathcal{Y}$. We refer to their joint probability density function (PDF) of $\bm{X}$ as $f_{\bm{X}}$. $Y$ is the random model response based on which we define if the system fails. 
This is done by defining the limit-state function $g(\bm{x})$ that by convention assumes values $\leq 0$ if the system fails.
We call $ \mathrm{F} = \{g(\bm{x}) \leq 0\}$ the failure event and the collection of all input values that result in system failure the failure domain $ \{\bm{x} : g(\bm{x}) \leq 0\}$.  The system's probability of failure is given as the integral of the input joint PDF over the failure domain:
\begin{equation}
	\label{eq:pf}
	\mathbb{P}(\mathrm{F}) = \int_{\mathcal{X}} \mathrm{I}[g(\bm{x}) \leq 0]  f_{\bm{X}}(\bm{x}) \mathrm{d}\bm{x}.
\end{equation}
$\mathrm{I}[\cdot]$ is an indicator function that equals $1$ on the set defined in its argument and $0$ outside that set. \cref{eq:pf} is usually difficult to compute accurately due to the rarity of the failure event ($\mathbb{P}(\mathrm{F})$ is small). Tools specifically designed to estimate the probability of failure are referred to as structural reliability methods \cite{Ditlevsen1996, Lemaire2009} (SRM) or rare event probability estimation methods.
\\~\\
Among these methods, one can distinguish approximation-based methods (e.g., the first- and second-order reliability method as discussed in \cite{Rackwitz1978,DerKiureghian2005}) that approximate the LSF and compute the failure probability associated with the approximation, and sampling-based methods that aim at reducing the variance of the crude Monte Carlo (MC) estimator. Modern sampling-based methods are often sequential MC-type methods  such as sequential importance sampling (SIS) \cite{Beaurepaire2013,Papaioannou2016}, subset simulation (SUS) \cite{Au2001} or the cross-entropy-based importance sampling method (CE-IS) \cite{Papaioannou2019b, Kroese2013,Kurtz2013,Wang2016,Rubinstein2017}. Line-sampling \cite{Hohenbichler1988,Koutsourelakis2004,Papaioannou2021} combines aspects of sampling-based and approximation methods. \cite{Uribe2020} introduces a dimensionality reduction approach for CE-IS that helps combat the curse of dimensionality in high-dimensional reliability problems. For the sake of efficiency, SRM are frequently paired with surrogate models that emulate $g$ at a fraction of the original computational cost. To this end, active learning procedures are often used to train Gaussian process regression surrogate models in the vicinity of the failure hypersurface $g=0$ \cite{Bichon2008, Echard2011}. While Gaussian process models are used most frequently for active learning-adapted surrogate models in SRM, other surrogate types such as polynomial chaos expansions have been used successfully in a similar fashion \cite{Marelli2018}. Recently, \cite{Ehre2022} introduced an approach to combine dimensionality reduction with active-learning-adapted partial least squares-based polynomial chaos expansion models.
\\~\\
In the analysis and design of engineering systems under uncertainty, it is often equally important to estimate both $\mathbb{P}(\mathrm{F})$ and the relative influence each of the uncertain parameters has on $\mathbb{P}(\mathrm{F})$, i.e.,  sensitivity measures of $\mathbb{P}(\mathrm{F})$ with respect to the components of $\bm{X}$. This is called reliability sensitivity analysis (RSA). 
If the interest is in sensitivity measures of the components of $\bm{X}$ with respect to more general model output $Y$, one speaks of sensitivity analysis of model output (SAMO).
Conceptually, RSA methods can be considered a subset of SAMO. In practice, owing to the rarity of the failure event that is targeted in the RSA setting, dedicated methods are usually required in order to accurately estimate sensititvies in the reliability setting.
Sensitivity measures (reliability sensitivity measures are no exception here) may be discerened based on whether they are computed with respect to random inputs $\bm{X}$ or deterministic parameters of either $g$ or $f_{\bm{X}}$. Further, one may discern local and global approaches, where the most common local sensitivity measure is the partial derivative of $\mathbb{P}(\mathrm{F})$ with respect to the parameters/inputs of interest \cite{Rubinstein1986,Wu1994,Au2005,Lu2008,Song2009,Papaioannou2018}. 
\cite{Wang2013b} proposed to average local failure probability derivatives to obtain a derivative-based global reliability sensitivity measure.
In the context of approximative SRM such as the first- and second-order reliability methpds (FORM, SORM), the directional cosines of the most probable point of failure provide a reliability sensitivity measure known as $\alpha$-factors  \cite{Hohenbichler1986,DerKiureghian2005}. 
Recently,  \cite{Papaioannou2021b} showed that $\alpha$-factors may be interpreted as global, variance-based sensitivity measures of the LSF approximation produced by FORM. 
Variance-based sensitivity indices are among the most frequently used SA measures and appear in the context of RSA in different forms:  \cite{Wei2012} proposed to compute a variance decomposition of the indicator function of the LSF in order to determine which input is responsible for what percentage of the total variance of the indicator function. \cite{Perrin2019,Li2019} suggested an efficient approach for computing first-order variance-based indices of the indicator function using samples from the failure domain and kernel density estimation. \cite{Morio2011,Wang2013} both compute the variance-based sensitivity indices of the failure probability conditional on a set of input distribution parameters with respect to said parameters using importance sampling and active-learning-based Gaussian process surrogates, respectively.
There exists a variety of other global sensitivity analysis methods that has been utilized in the RSA context, including moment-independent measures \cite{Cui2010}, Shapley values \cite{Ilidrissi2021} and the expected value of (partial) perfect information \cite{Straub2022}. 
\\~\\
Variance-based sensitivity measures are most often applied to problems with statistically independent inputs. This is because in the case of dependent inputs, variance-based indices cannot be interpreted in terms of a decomposition of the model output variance: in this case, variance-bases sensitivity indices will contain covariance contributions stemming from the variable dependence that preclude a unique interpretation of the indices. In the more general context of SAMO (without focus on rare events), it has been proposed to evaluate variance-based sensitivity indices that isolate the contribution of each variable excluding its dependence with other variables \cite{Mara2012}. In this paper, we apply these indices in the context of RSA and propose a method for their efficient estimation. In particular, we extend on the failure sample-based approach to variance-based RSA in \cite{Li2019}. Our method can be applied with any sampling-based reliability method that produces failure samples and does not require additional LSF evaluations.  We introduce the basic concepts of variance-based sensitivity analysis for model output and reliability in \cref{sec:rsa} and review extensions of these concepts to dependent inputs in \cref{sec:dependency}. In \cref{sec:trafos}, we introduce hierarchical transformations, which are a cornerstone of our method that we then describe in \cref{sec:method}. In \cref{sec:examples}, we demonstrate the performance of the proposed method in several numerical examples using MC, CE-IS and SUS. We provide some concluding remarks in \cref{sec:concluding_remarks}.
\section{Variance-based reliability sensitivity analysis}
\label{sec:rsa}
An important global sensitivity measure targeting random parameters is based on the variance decomposition of the model $\mathcal{Y}$. Any square-integrable function $\mathcal{Y}$ can be expressed as
 \begin{equation}
 	\label{eq:sobol_decomp}
 	\mathcal{Y}(\bm{X}) =  \hspace{-.5cm} \sum_{\bm{v} \in \mathcal{P}(\{1,\dots,d\})} \hspace{-.5cm} \mathcal{Y}_{\bm{v}}(X_{\bm{v}}),
 \end{equation}
where $\mathcal{P}$ denotes the power set and for the empty set we define $\mathcal{Y}_{\bm{v}} = \mathcal{Y}_{\emptyset}$, which is a constant. Any other $\mathcal{Y}_{\bm{v}}$ can be regarded as a function depending on exactly $\bm{X}_{\bm{v}} = \{X_j\}_{j \in \bm{v}}$.
The expansion in \cref{eq:sobol_decomp} exists and is unique if $\bm{X}$ is a vector of independent random variables such that $f_{\bm{X}}(\bm{x}) = \prod_{i=1}^d f_{X_i}(x_i)$ and if the expectation of any $\{\mathcal{Y}_{\bm{v}}(\bm{X}_{\bm{v}})\}_{\bm{v} \in \mathcal{P}(\{1,\dots,d\})\textbackslash \emptyset}$ with respect to any component of $\bm{X}_{\bm{v}}$ equals zero \cite{Sobol1993}, i.e., 
\begin{equation}\label{eq:zeromarginalmean}
	\int \mathcal{Y}_{\bm{v}}(\bm{X}_{\bm{v}}) f_{X_i}(x_i) \mathrm{d}x_i = 0 ~ \forall ~i \in \bm{v}.
\end{equation}
Consequences of the above constraint are $\mathcal{Y}_\emptyset = \mathbb{E}[\mathcal{Y}]$ and that the $\{\mathcal{Y}_{\bm{v}}(\bm{X}_{\bm{v}})\}_{\bm{v} \in \mathcal{P}(\{1,\dots,d\}) \mathrm{\textbackslash} \emptyset}$ are pairwise orthogonal:
\begin{equation}\label{eq:orthogonality}
	\mathbb{E}[\mathcal{Y}_{\bm{v}}(\bm{X}_{\bm{v}})\mathcal{Y}_{\bm{w}}(\bm{X}_{\bm{w}})] = 0~\text{iff}~\bm{v} \neq \bm{w}.
\end{equation}  
The resulting expansion is known as the Sobol'-Hoeffding-decomposition. From \cref{eq:orthogonality} follows that the variance of $\mathcal{Y}$ is the sum of the partial variances of the decomposition terms:
 \begin{equation}\label{eq:variance_decomp}
	\mathbb{V}[\mathcal{Y}] = \hspace{-.5cm} \sum_{\bm{v} \in \mathcal{P}(\{1,\dots,d\})} \hspace{-.5cm} \mathbb{V}[\mathcal{Y}_{\bm{v}}]. 
 \end{equation}
Collecting terms on the right-hand side of \cref{eq:variance_decomp} that relate to a particular subset of inputs $\bm{X}_{\bm{v}}$ and dividing by the total variance of $\mathcal{Y}$, we obtain the Sobol' index \cite{Sobol1993} $S_{\bm{v}}$ and the total  Sobol' index \cite{Homma1996} $S^T_{\bm{v}}$ of said subset as
\begin{equation}\label{eq:indices_1}
	S_{\bm{v}} \vcentcolon = \frac{\mathbb{V}[\mathcal{Y}_{\bm{v}}]}{\mathbb{V}[\mathcal{Y}]},~~S^T_{\bm{v}} \vcentcolon =  \frac{\mathbb{V}[\sum_{\bm{v} \subseteq \bm{w}}\mathcal{Y}_{\bm{w}}]}{\mathbb{V}[\mathcal{Y}]}.
\end{equation}
The Sobol' index thus measures the variance fraction contributed to $\mathbb{V}[\mathcal{Y}]$ by $\bm{X}_{\bm{v}}$ only through $\mathcal{Y}_{\bm{v}}$. Conversely, the total Sobol' index also includes the variance fractions contributed by the interaction of $\bm{X}_{\bm{v}}$ with any other inputs.
Using \cref{eq:zeromarginalmean}, the terms of the Sobol'-Hoeffding decomposition can be computed recursively as
 \begin{equation}\label{eq:partial_shd_function}
 	\mathcal{Y}_{\bm{v}}(\bm{X}_{\bm{v}}) = \mathbb{E}[\mathcal{Y}(\bm{X})|\bm{X}_{\bm{v}}] - \sum_{\bm{w} \subset \bm{v}} \mathbb{E}[\mathcal{Y}(\bm{X})|\bm{X}_{\bm{w}}].
 \end{equation}
Focussing on first-order indices ($|\bm{v}| = 1$) and plugging \cref{eq:partial_shd_function} in \cref{eq:indices_1}, we obtain
\begin{equation}\label{eq:indices_2}
	S_{i} = \frac{\mathbb{V}[\mathbb{E}[\mathcal{Y}(\bm{X})|X_{i}] ]}{\mathbb{V}[\mathcal{Y}]},~~S^T_{i} =  1 -   \frac{\mathbb{V}[\mathbb{E}[\mathcal{Y}(\bm{X})|\bm{X}_{\sim i}] ]}{\mathbb{V}[\mathcal{Y}]}.
\end{equation}
These expressions are more convenient to construct estimators of $S_{i},S^T_{i}$ as often the variance decomposition of $\mathcal{Y}$ is difficult to identify. The naive approach to estimating the nested operators in \cref{eq:indices_2} is a double loop MC simulation. However, there exists a class of more efficient estimators -- so-called pick-freeze estimators -- that avoid this double loop \cite{Jansen1999,Saltelli2000,Saltelli2010}. 
\\~\\
\cite{Wei2012}  proposed computing Sobol' indices of $\mathcal{Y}(\bm{X}) = \mathrm{I}[g(\bm{X}) \leq 0]$.
As the indicator function is Bernoulli-distributed with success parameter the probability of failure $\mathbb{P}(\mathrm{F})$, the convergence rate of pick-freeze estimators for this choice of $\mathcal{Y}$ is often slow to the point of being computationally unaffordable ($n$ grows very large for a fixed coefficient of variation (c.o.v.) of the pick-freeze estimators).
The Sobol' indices of the indicator function may be rewritten as \cite{Li2012}
\begin{equation}\label{eq:ROSobol}
	S_i 
	=  \frac{\mathbb{V}[\mathbb{P}(\mathrm{F}|X_i)]}{\mathbb{V}[\mathrm{I}[g \leq 0]]},~~S^T_{i} =  1 -   \frac{\mathbb{V}[\mathbb{P}(\mathrm{F}|\bm{X}_{\sim i})]}{\mathbb{V}[\mathrm{I}[g \leq 0]]}.
\end{equation}
\cite{Perrin2019} and \cite{Li2019} independently applied Bayes' rule to express the conditional failure probability in \cref{eq:ROSobol} as 
\begin{equation}\label{eq:pf_cond}
 \mathbb{P}(\mathrm{F}|X_i)	= \frac{ f_{X_i|\mathrm{F}}(x_i)\mathbb{P}(\mathrm{F})}{ f_{X_i}(x_i)}.
\end{equation}
Based on \cref{eq:pf_cond}, $\mathbb{P}(\mathrm{F}|X_i)$ may be computed efficiently by approximating $ f_{X_i|\mathrm{F}}(x_i)$ with a kernel density estimator using a set of failure samples that follow $ f_{X_i|\mathrm{F}}(x_i)$. A set of failure samples can be obtained as a byproduct of computing $\mathbb{P}(\mathrm{F})$ with any sampling-based reliability method such as MC, importance sampling, SUS \cite{Au2001} or the sequential and cross-entropy importance sampling varieties referenced in \cref{sec:intro}. In this way, first-order reliability Sobol' indices may be computed as byproduct of a single run of a rare event estimation method. Computing total Sobol' indices of the indicator with this approach requires $d-1$-dimensional kernel density estimates and is thus limited to low-dimensional problems \cite{Li2019}. 
\section{Dependent input measures}
\label{sec:dependency}
When the inputs $\bm{X}$ are no longer independent, it is still possible to construct the decomposition in \cref{eq:sobol_decomp} using summands that satisfy \cref{eq:zeromarginalmean}. However, this decomposition will not be unique \cite{Chastaing2012} and \cref{eq:zeromarginalmean} no longer implies orthogonality amongst the summands as in \cref{eq:orthogonality}. Hence, the variance of $\mathcal{Y}$ is no longer the sum of partial variances of the $\{\mathcal{Y}_{\bm{v}}(\bm{X}_{\bm{v}})\}_{\bm{v} \in \mathcal{P}(\{1,\dots,d\})\textbackslash \emptyset}$ but also contains covariances of pairs of decomposition terms. 
\cite{Xu2008} compute the total variance contribution associated with any $X_i$ using different univariate linear regressions. They use a second regression on orthogonalized adaptations of the $X_i$ to determine their uncorrelated effects. The former correspond to partial variances and covariances and measure the contribution of any $X_i$ to $\mathbb{V}[Y]$ through the computational model $\mathcal{Y}$ as well as the probabilistic model $f_{\bm{X}}$.  Conversely, the latter includes no partial covariances and hence measures the uncorrelated contribution of any $X_i$ on $\mathbb{V}[Y]$ through $\mathcal{Y}$ only (i.e., as if $\bm{X}$ were an uncorrelated random vector). \cite{LiRabitz2010} also use a regression approach to distinguish between total and uncorrelated variance contributions but use a set of nonlinear features in the regression approach (B-splines). \cite{Caniou2012} fit a polynomial chaos expansion to $\mathcal{Y}$ assuming independent inputs but then use dependent samples to determine the different variance and covariance contributions. 
\cite{Kucherenko2012} suggested a modified pick-freeze estimator for the case of dependent inputs that does not require fitting a surrogate model, however, will not separate the influence of variable dependence.
\cite{Bedford1998} proposed computing variance-based sensitivities under dependent inputs by orthogonalizing conditional expectations of the output. Using Gram-Schmidt, they transform the dependent input to an uncorrelated random vector (not necessarily independent). However, for $d$ inputs, there are $d!$ orderings in which inputs can be orthogonalized, which produces ambiguity in the variance-based sensitivity measures obtained in this way: If the variance contribution of a transformed variable strongly depends on the order in which inputs are orthogonalized, it is unclear which ordering to trust.
\cite{Mara2012} picked up on the idea presented in \cite{Bedford1998} and suggested an approach to resolve the interpretation issue that comes with the non-uniqueness of the high-dimensional model representation in case of dependent variables. They repeat the orthogonalization procedure for $d$ different orderings. 
This idea extends to more general variable transformations than orthogonal conditional expectations and has been investigated in \cite{Mara2016} using the Rosenblatt transformation \cite{Rosenblatt1952} as well as a Spearman rank correlation-based approach that is due to \cite{ImanConover1982}. In the next section, we detail the Rosenblatt transform and elaborate on its hierarchical structure, which will prove central to our approach.
Further, we discuss a special case that arises from applying the Rosenblatt transform to random vectors with a Gaussian copula:  the Nataf model \cite{Nataf1962}.
 \section{Hierarchical transformations of dependent inputs}
 \label{sec:trafos}
 Let $\bm{U}: \Omega \rightarrow \mathcal{U} \subseteq \mathbb{R}^d$ and $\bm{X}: \Omega \rightarrow \mathcal{X} \subseteq \mathbb{R}^d$ be continuous, real-valued $d$-dimensional random vectors.
 We seek a transformation $\bm{U} = T(\bm{X})$ such that the components of $\bm{U}$ are independent and marginally distributed as some standard distribution type.
Conventionally, the independent standard-normal space is used as the image space of $T$ in the structural reliability literature. This is to say $\bm{U} \sim \varphi_d(\bm{u})$ is an independent standard-normal random vector. Thus, $\varphi_d(u) = \prod_{i=1}^d \varphi(u_i)$. For the PDF and cumulative distribution function (CDF) of a standard-normal random variable we write $\varphi(\cdot)$ and $\Phi(\cdot)$, respectively.
Next, we introduce the Rosenblatt transformation, which is a transformation $\bm{U} = T(\bm{X})$ possessing a specific hierarchical structure:
\begin{definition}[Rosenblatt transformation]
	\label{def:rosenblatt}
	Let $\bm{U}: \Omega \rightarrow \mathbb{R}^d$ be an independent standard-normal random vector and $\bm{X}: \Omega \rightarrow \mathcal{X} \subseteq \mathbb{R}^d$ be a continuous, real-valued $d$-dimensional random vector with marginal CDF $F_{X_1}$ and conditional CDFs $F_{X_1|X_2}, F_{X_3|X_1,X_2},\dots, F_{X_d|X_1,\dots,X_{d-1}}$. Then, the Rosenblatt transformation from $\bm{X}$ to $\bm{U}$ is given as
	$$
	\begin{bmatrix} U_1 \\ U_{2}  \\ \vdots \\ U_{d}  \end{bmatrix} = \begin{bmatrix} \Phi^{-1}  \circ F_{X_1}(X_1) \\ \Phi^{-1}  \circ F_{X_2|X_1}(X_2|X_1) \\ \vdots \\ \Phi^{-1}  \circ F_{X_d|X_1,\dots, X_{d-1}}(X_d|X_1,\dots,X_{d-1})]\end{bmatrix}.
	$$
\end{definition}
	In practice, it is often difficult to compute all conditional CDFs required in constructing the Rosenblatt transformation following \cref{def:rosenblatt}. Thus, we next consider a model for the joint distribution that renders conditional CDFs tractable -- the so-called Nataf model \cite{Nataf1962}:
\begin{definition}[Nataf model]
	\label{eq:nataf}
	Let $\bm{X}: \Omega \rightarrow \mathcal{X} \subseteq \mathbb{R}^d$ be a continuous, real-valued $d$-dimensional random vector with marginal CDFs $\{F_{X_i}\}_{i=1}^d$ and covariance matrix $\bm{\Sigma} = \mathbb{E}[(\bm{X}-\mathbb{E}[\bm{X}])(\bm{X}-\mathbb{E}[\bm{X}])^{\mathrm{T}}]$. 
	Further, let $\bm{Z}: \Omega \rightarrow \mathbb{R}^d$ with components $Z_i = \Phi^{-1} ( F_{X_i}(X_i))$ ($i=1,\dots, d$) be a standard-normal random vector with covariance matrix $\bm{\Sigma}_{\bm{Z}}$.
	Then, the Nataf model of the joint PDF of $\bm{X}$ is given as
	\begin{equation*}
		f_{\bm{X}}(\bm{x}) = \frac{f_{X_1}(x_1) \cdots f_{X_d}(x_d)}{\varphi(z_1) \cdots \varphi(z_d)} \varphi_d(\bm{z},\bm{\Sigma}_{\bm{Z}}),
	\end{equation*}
where $\varphi_d(\bm{z},\bm{\Sigma}_{\bm{Z}})$ is the $d$-dimensional standard-normal PDF with covariance matrix $\bm{\Sigma}_{\bm{Z}}$ and the entries of the covariance matrix $\bm{\Sigma}_{\bm{Z}}$ are implicitly related to their corresponding entries in $\bm{\Sigma}$ via 
\begin{equation}
	\label{eq:sigma_z}
\begin{aligned}
		&[\bm{\Sigma}]_{ij} =\int\limits_{-\infty}^{\infty} \int\limits_{-\infty}^{\infty} F_{X_i}^{-1} \circ \Phi(z_i)  \cdot F_{X_j}^{-1} \circ \Phi(z_j) \\ 
		&\cdot \varphi_2\left(\begin{bmatrix} z_i \\ z_j \end{bmatrix}, \begin{bmatrix} 1 & \!\!\!\!\!\! [\bm{\Sigma}_{\bm{Z}}]_{ij} \\  [\bm{\Sigma}_{\bm{Z}}]_{ij} & \!\!\!\!\!\!1 \end{bmatrix}\right) \mathrm{d}z_i \mathrm{d}z_j - \mathbb{E}[X_i] \mathbb{E}[X_j].
	\end{aligned}
\end{equation}
\end{definition}
Note, that the Nataf model only requires knowledge of the marginal distributions of each input parameter as well as their covariance structure. Higher-order dependencies are not captured.
The Nataf model can also be understood as a Gaussian copula model \cite{Lebrun2009} acting on $[F_{X_1}(x_1),\dots,F_{X_d}(x_d)]{}$, i.e.,  the marginal transformation of $\bm{X}$ to the d-dimensional standard uniform distribution.
\\~\\
Using the Rosenblatt transformation to transform the dependent Gaussian random vector $\bm{Z}$ into its independent counterpart $\bm{U}$, we achieve a hierarchical transformation based on the Nataf model. 
\begin{definition}[Rosenblatt transform for the Nataf model]
Let $\bm{U}: \Omega \rightarrow \mathbb{R}^d$ be an independent standard-normal random vector and $\bm{X}: \Omega \rightarrow \mathcal{X} \subseteq \mathbb{R}^d$ be a continuous, real-valued $d$-dimensional random vector with marginal CDFs $\{F_{X_i}\}_{i=1}^d$ and covariance matrix $\bm{\Sigma}$. Further, let $\mathbf{A}$ be the Cholesky factor of $\bm{\Sigma}_{\bm{Z}} $ such that $\bm{\Sigma}_{\bm{Z}} = \mathbf{A}\mathbf{A}^{\mathrm{T}}$ and $\mathbf{A}$ is lower-triangular.
	Then, the Rosenblatt transformation from $\bm{X}$ to $\bm{U}$ is given as \cite{Lebrun2009c}
	$$\begin{bmatrix} U_1 \\ U_{2}  \\ \vdots \\ U_{d}  \end{bmatrix} = \mathbf{A}^{-1} \begin{bmatrix} \Phi^{-1}  \circ F_{X_1}(x_1) \\ \Phi^{-1}  \circ F_{X_2}(x_2) \\ \vdots \\ \Phi^{-1}  \circ F_{X_d}(x_d)]\end{bmatrix} .
	$$
\end{definition}
The advantage of applying the Rosenblatt transform to the Nataf model instead of general joint distributions (as in \cref{def:rosenblatt}) is due to the fact that all conditional CDFs of the correlated Gaussian random vector $\bm{Z}$ are tractable and the transformation can be readily computed using the Cholesky decomposition of $\bm{\Sigma}_{\bm{Z}}$. It is of particular relevance to the approach presented in this paper that the Rosenblatt transform of the Nataf model is readily available for any ordering of the components of $\bm{X}$.
\section{Failure sample-based method}
\label{sec:method}
\begin{figure*}[!ht]
	\centering
	\includegraphics[width=\textwidth]{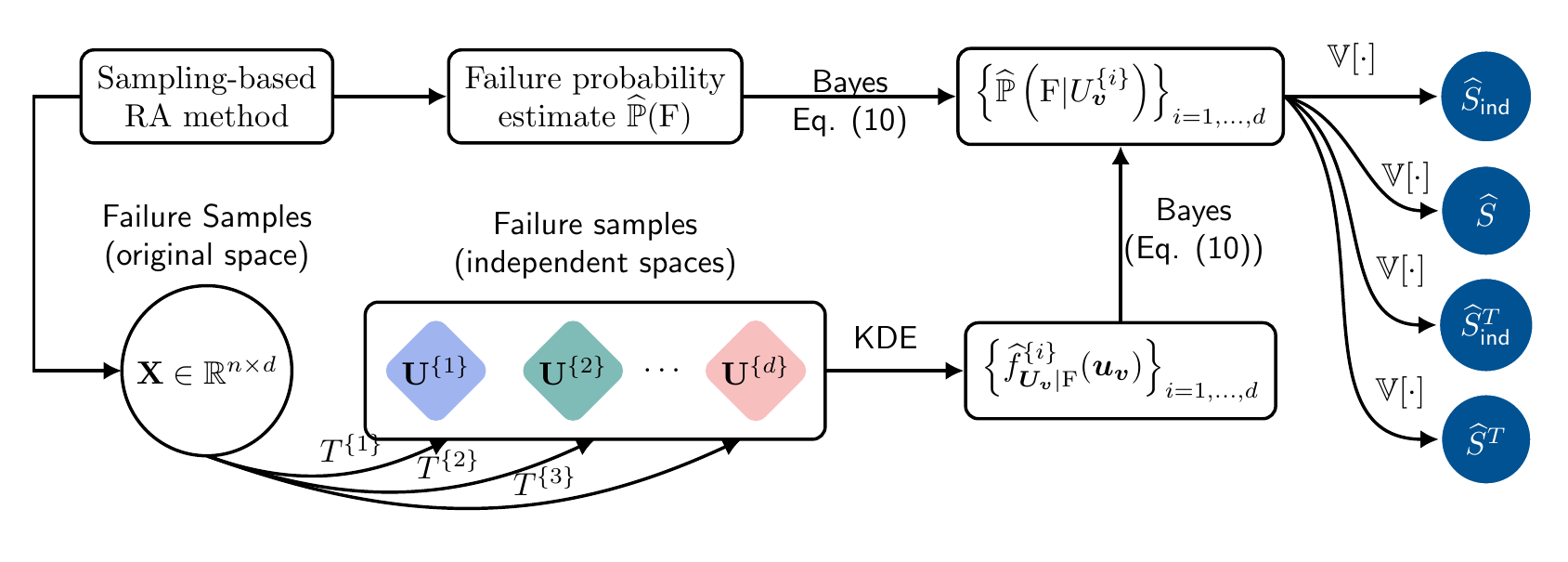}
	\caption{Computational graph of estimating the full set of variance-based reliability sensitvitiy indices for models with dependent inputs using a single RA run.}
	\label{fig:flow_diag}
\end{figure*}

The benefit of using hierarchical transformations is that the influence of each $X_i$ can be clearly isolated depending on its position in the transformation; in \cref{def:rosenblatt}, the full contribution of $X_1$ is retained in $U_1$ through the marginal transformation. On the other hand, $X_d$ is transformed to $U_d$ conditional on $X_1,\dots,X_{d-1}$ whereby only the independent contribution of $X_d$ is retained in $U_d$. There are $d!$ possibilities to order $\{X_1,\dots,X_d\}$ in the hierarchical transformation. However, as pointed out in \cite{Mara2012,Mara2016}, due to the hierarchical structure of the transformation, it is only necessary to consider $d$ transformations to isolate the independent contribution of each input from its total effect in terms of the output variance. 
 \begin{figure}[!ht]
	\label{fig:permutations}
	\centering
	\includegraphics[width=.4\textwidth]{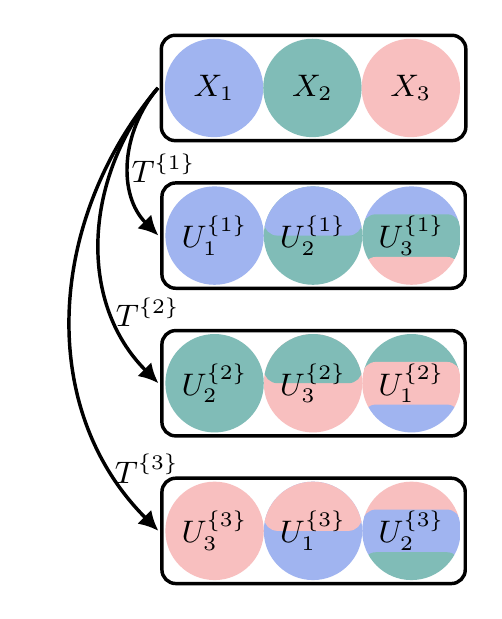}
	\caption{Illustration of the collection of the $d$ hierarchical transformations considered in this paper for the case $d=3$ (colors indicate dependence on $\bm{X}$-components): $T^{\{1\}}: \bm{X} \to \bm{U}^{\{1\}}$, marginally transforms $X_1$, then transforms $X_2$ conditional on $X_1$ and then $X_3$ conditional on $X_1, X_2$. For $T^{\{2\}}: \bm{X} \to \bm{U}^{\{2\}}$ and $T^{\{3\}}: \bm{X} \to \bm{U}^{\{3\}}$, the procedure is repeated with the components of $\bm{X}$ cyclically shifted to the left by one and two positions, respectively. Colors indicate on which $\bm{X}$-components any $U_j^{\{j\}}$ depends.}
\end{figure}
More precisely, it is sufficient, to consider all $d$ cyclic shifts of the ordered set $\{X_1,\dots,X_d\}$ and define the set of hierarchical transformations \smash{$\{T^{\{i\}} \}_{i=1}^d$} with
\begin{equation}
\label{eq:trafos}
T^{\{i\}}: \begin{bmatrix} U_i^{\{i\}} \\ U_{i+1}^{\{i\}}  \\ \vdots \\  U_d^{\{i\}} \\ U_{1}^{\{i\}} \\ \vdots \\ U_{i-1}^{\{i\}}  \end{bmatrix} = \begin{bmatrix} h_1^{\{i\}}(X_i)\\ h_2^{\{i\}}(X_i,X_{i+1})\\ \vdots \\  h_{d-i+1}^{\{i\}}(X_i,\dots,X_d)\\ h_{d-i+2}^{\{i\}}(X_i,\dots,X_d,X_1) \\ \vdots \\  h_d^{\{i\}}(X_i,\dots,X_d,X_1,\dots,X_{i-1}) \end{bmatrix},
\end{equation}
where for the last component of $T^{\{1\}}$, we set $X_0 \vcentcolon = X_d$ and the precise form of any $h_j^{\{i\}}$ depends on whether we use a generic joint distribution or the Nataf model. \cref{fig:permutations} illustrates variable dependencies in each transformation arising from a cyclic shift in the case of $d=3$.
For the generic case, obtaining all $h_j^{\{i\}}$ corresponds to working out all necessary conditional CDFs associated to each of the $d$ variable orderings. For the Nataf model case, it is sufficient to determine $\bm{\Sigma}_{\bm{U}}$ once and then compute the Cholesky factor of $\bm{\Sigma}_{\bm{U}}$ with rows and columns permuted according to each of the $d$ variable orderings.
$T^{\{i\}}$ is based on the variable ordering $\{X_i,X_{i+1}\dots,X_{i-1}\}$, which corresponds to $i-1$ cyclic left shifts of $\{X_1,\dots,X_d\}$.
Based on the consideration that, under $T^{\{i\}}$, $U_i^{\{i\}}$ represents the full effect of $X_i$ whereas  $U_{i-1}^{\{i\}}$ represents only the independent effect of $X_{i-1}$, one defines
\begin{equation}
\begin{aligned}
S_{i,\text{ind}} &= \frac{\mathbb{V}[\mathbb{E}[\mathcal{Y}(\bm{X})|U_{i}^{\{i+1\}}]]}{\mathbb{V}[\mathcal{Y}(\bm{X})]},&
S_{i} &= \frac{\mathbb{V}[\mathbb{E}[\mathcal{Y}(\bm{X})|U_i^{\{i\}}]]}{\mathbb{V}[\mathcal{Y}(\bm{X})]}\\
S^T_{i,\text{ind}} &=  1 -   \frac{\mathbb{V}[\mathbb{E}[\mathcal{Y}(\bm{X})|\bm{U}^{\{i\}}_{\sim i}]]}{\mathbb{V}[\mathcal{Y}(\bm{X})]},&S^T_{i} &=  1 -   \frac{\mathbb{V}[\mathbb{E}[\mathcal{Y}(\bm{X})|\bm{U}^{\{i+1\}}_{\sim i}]]}{\mathbb{V}[\mathcal{Y}(\bm{X})]}. 
\end{aligned}
\end{equation}
Here, $S_i$ is the standard Sobol' index measuring the combined effect of $X_i$ via both probabilistic and computational model, whereas $S_{i,\text{ind}}$ is an independent counterpart of $S_i$ measuring only the effect of $X_i$ via the computational model (as if it were independent of the other input parameters). Likewise, an independent total Sobol' index $S^T_{i,\text{ind}}$ is defined, that is computed based on $\bm{U}^{\{i\}}$ rather than $\bm{U}^{\{i+1\}}$, however, as the conditional expectation is with respect to the set $\bm{v} = \sim i$.
\cite{Mara2016} use pick-freeze estimators in $d$ repeated SAMO runs to obtain all independent and full Sobol'/total Sobol' indices. With $n$ independent samples per SAMO run, this requires a total of $0.5 n d(d+2)$ model calls \cite{Saltelli2010}.
\\~\\
As explained in \cref{sec:rsa}, pick-freeze estimators are inefficient for RSA. Instead, we may use \cref{eq:ROSobol}, \cref{eq:pf_cond} and the fact that $\mathbb{V} \left[ \mathrm{I}[ g(\bm{X}) <0] \right] = \mathbb{P}(\mathrm{F})(1-\mathbb{P}(\mathrm{F})) $ to write 
\begin{align}
	\label{eq:indices_3}
	S_{i} &=  \frac{\mathbb{P}(\mathrm{F})}{1 - \mathbb{P}(\mathrm{F})} \mathbb{V} \left[\frac{ f_{X_i|\mathrm{F}}(X_i)}{ f_{X_i}(X_i)}  \right]  \\
	S^T_{i} &= 1 - \frac{\mathbb{P}(\mathrm{F})}{1 - \mathbb{P}(\mathrm{F})}\mathbb{V} \left[\frac{ f_{\bm{X}_{\sim i}|\mathrm{F}}(\bm{X}_{\sim i})}{ f_{\bm{X}_{\sim i}}(\bm{X}_{\sim i})}  \right] .
\end{align}
One may use any sample-based SRM to generate a single set of failure samples \smash{$\mathbf{X} \in \mathbb{R}^{n \times d}$} alongside the probability of failure estimate $\widehat{\mathbb{P}}(\mathrm{F})$. \cite{Li2019,Perrin2019} evaluate \cref{eq:indices_2} based on $\mathbf{X}$, which yields $\{S_i\}_{i=1}^d$ and $\{S^T_{\mathrm{ind},i}\}_{i=1}^d$. In order to obtain the entire set of  independent and full Sobol' and total Sobol' indices, we instead use the set of hierarchical transformations \smash{$\{T^{\{i\}} \}_{i=1}^d$} to generate $d$ different standard-normal sample sets \smash{$\{\mathbf{U}^{\{i\}} \}_{i=1}^d$}. We then estimate the required PDFs \smash{$f_{U_i|\mathrm{F}}, f_{U_{i-1}|\mathrm{F}}$} and \smash{ $f_{\bm{U}_{\sim i}|\mathrm{F}}, f_{\bm{U}_{\sim (i -1)}|\mathrm{F}}$} with a standard-normal kernel $\varphi(\bm{x})$ as
\begin{equation}
	\label{eq:kde}
	\widehat{f}^{\{i\}}_{\bm{U}_{\bm{v}}|\mathrm{F}}(\bm{u}_{\bm{v}}) = \frac{1}{n  |\mathbf{H}|^{\frac{1}{2}}}\sum\limits_{j=1}^n \varphi \left( \mathbf{H}^{-\frac{1}{2}} \left(\bm{u}_{\bm{v}} - \mathbf{U}^{\{i\}}_{\bm{v}}\right)\right),
\end{equation}
where $\bm{v} = \{i,i-1,\sim i, \sim (i-1)\}$. For the bandwidth covariance $\mathbf{H}$ we use a diagonal matrix and Silverman's rule to estimate the main diagonal elements as $\sqrt{H_{ii}} = (4/(|\bm{v}|+2))^{1/(|\bm{v}|+4)} \widehat{\sigma}_i$ with  $\widehat{\sigma}_i$ the empirical standard deviation of the sample column \smash{$\mathbf{U}^{\{i\}}_{v_i}$}.
Note that using Silverman's rule can yield inaccurate kernel density estimates for multimodal targets \cite[Chapter 4]{Haerdle1991}. This can be alleviated by using more sophisticated bandwidth selection methods such as maximum likelihood cross-validation \cite{Duin1976} or unbiased cross-validation \cite{Rudemo1982,Bowman1984}.
Plugging these density estimates and the probability of failure estimate in \cref{eq:indices_3}, we obtain estimators for the full set of variance-based reliability sensitivity indices in standard-normal space: 
\begin{align}\label{eq:indices_4}
	\widehat{S}_{i} & = \frac{\widehat{\mathbb{P}}(\mathrm{F})}{1-\widehat{\mathbb{P}}(\mathrm{F})} \mathbb{V} \left[\frac{ \widehat{f}^{\{i\}}_{U_i|\mathrm{F}}(U_i)}{ \varphi(U_i)}  \right]\\
	\widehat{S}_{i,\mathrm{ind}} &= \frac{\widehat{\mathbb{P}}(\mathrm{F})}{1-\widehat{\mathbb{P}}(\mathrm{F})} \mathbb{V} \left[\frac{ \widehat{f}^{\{i+1\}}_{U_i|\mathrm{F}}(U_i)}{ \varphi(U_i)}  \right] \\
	\widehat{S}^T_{i} &= 1 - \frac{\widehat{\mathbb{P}}(\mathrm{F})}{1 - \widehat{\mathbb{P}}(\mathrm{F})}\mathbb{V} \left[\frac{ \widehat{f}^{\{i+1\}}_{\bm{U}_{\sim i}|\mathrm{F}}(\bm{U}_{\sim i})}{ \varphi(\bm{U}_{\sim i})}  \right]\\
	\widehat{S}^T_{i,\mathrm{ind}} &= 1 - \frac{\widehat{\mathbb{P}}(\mathrm{F})}{1 - \widehat{\mathbb{P}}(\mathrm{F})}\mathbb{V} \left[\frac{ \widehat{f}^{\{i\}}_{\bm{U}_{\sim i}|\mathrm{F}}(\bm{U}_{\sim i})}{ \varphi(\bm{U}_{\sim i})}  \right].
\end{align}
\cref{fig:flow_diag} summarizes the computational steps involved in estimating the above indices.
In the next section, we showcase the FS estimator using different RA methods and benchmark these against two existing techniques to estimate the full set of variance-based sensitivity indices for model ouput in \cref{eq:indices_4}.

\label{sec:approach}
\section{Examples}
\label{sec:examples}
\subsection{Nonlinear test function}
\label{sec:nonlinfun}
\begin{figure}[!ht]
	\centering
	\subfloat{\includegraphics[width=.5\textwidth]{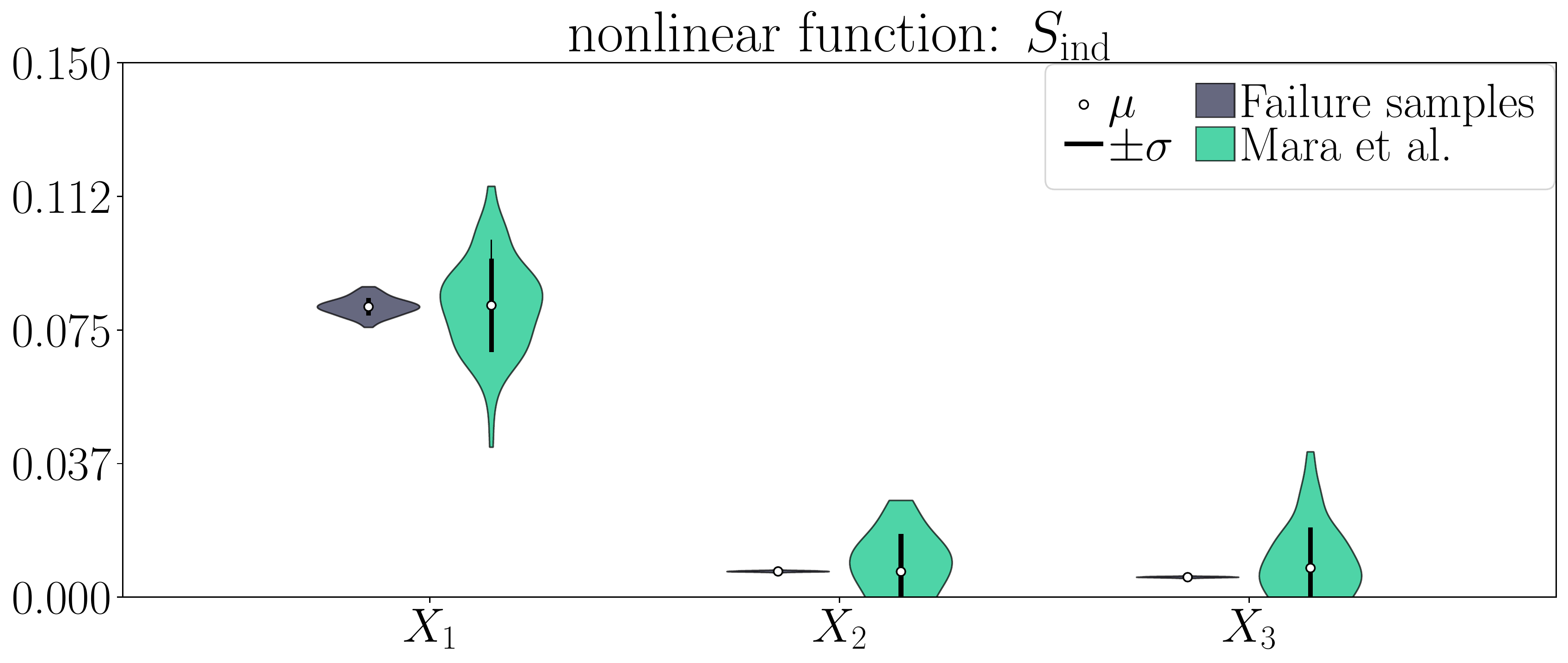}}\\
	\subfloat{\includegraphics[width=.5\textwidth]{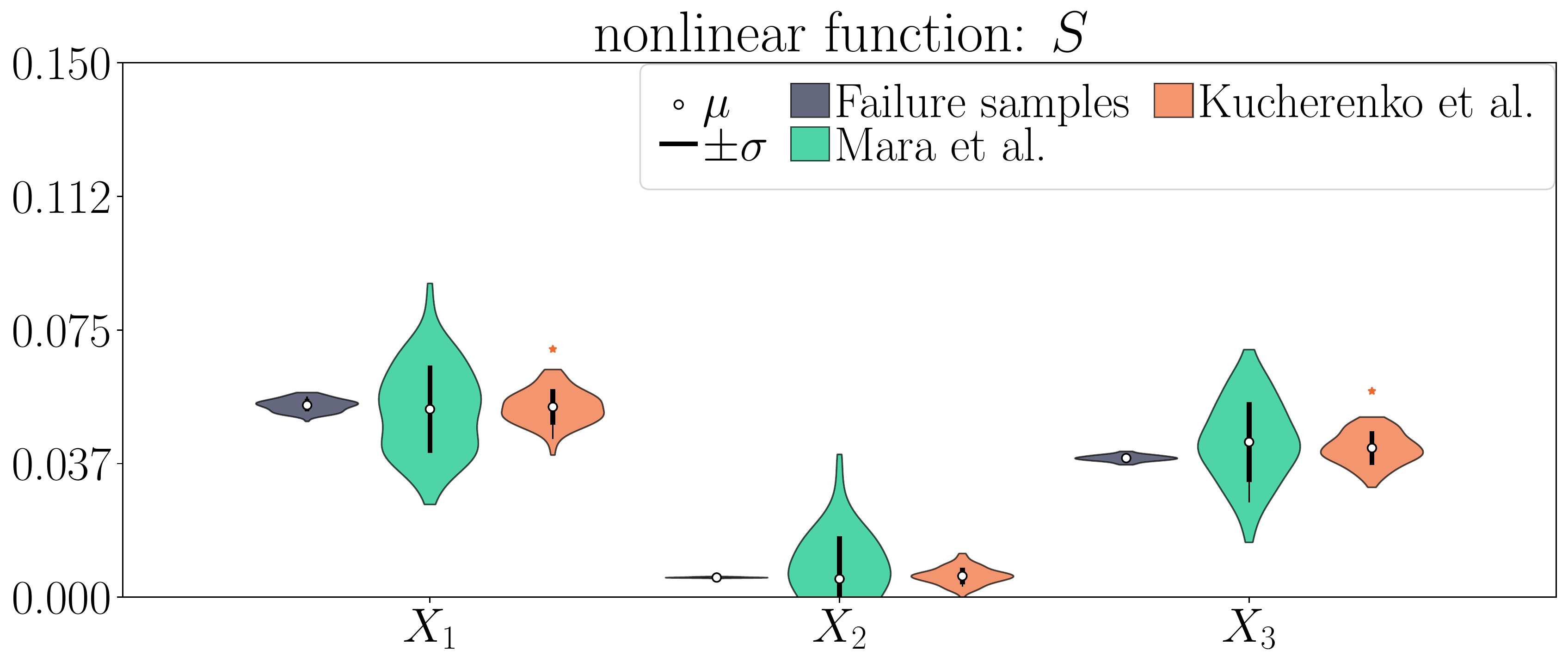}}\\
	\subfloat{\includegraphics[width=.5\textwidth]{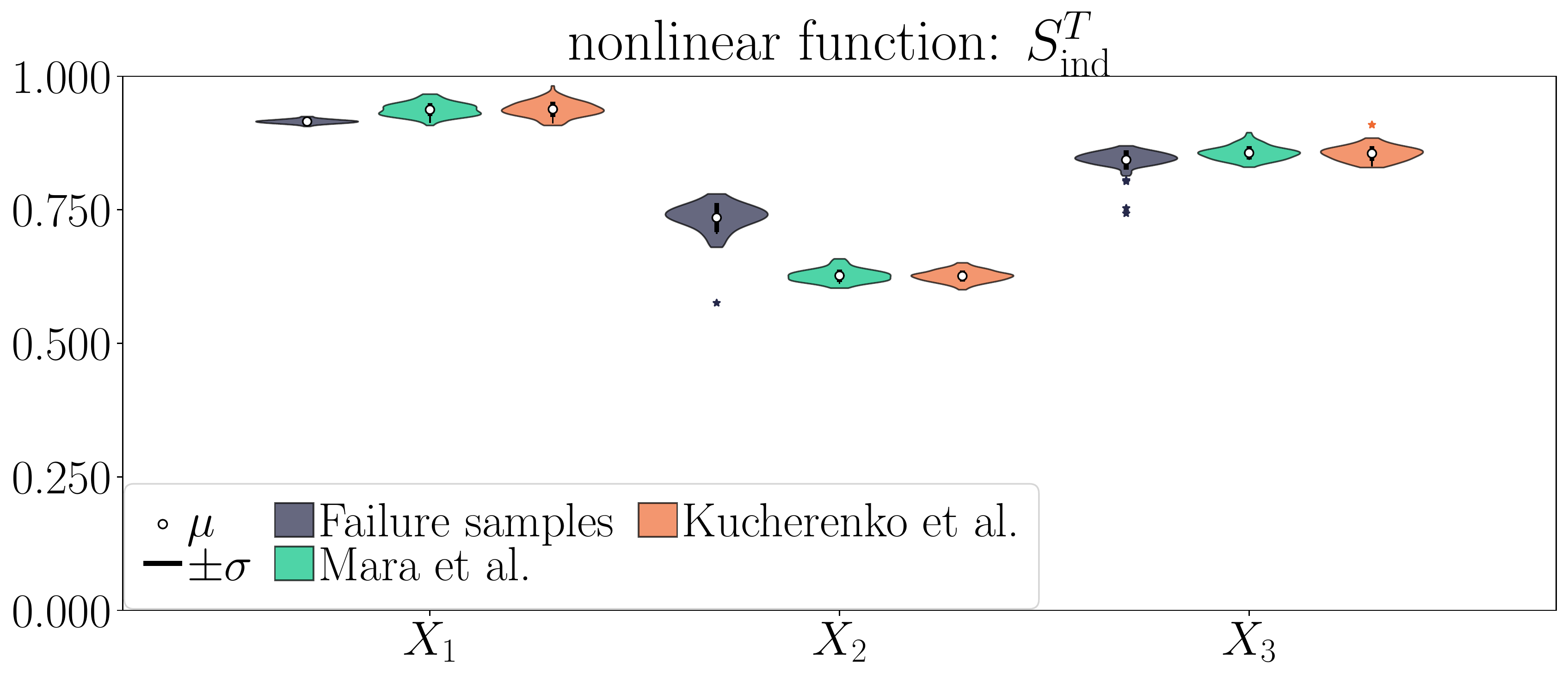}}\\
	\subfloat{\includegraphics[width=.5\textwidth]{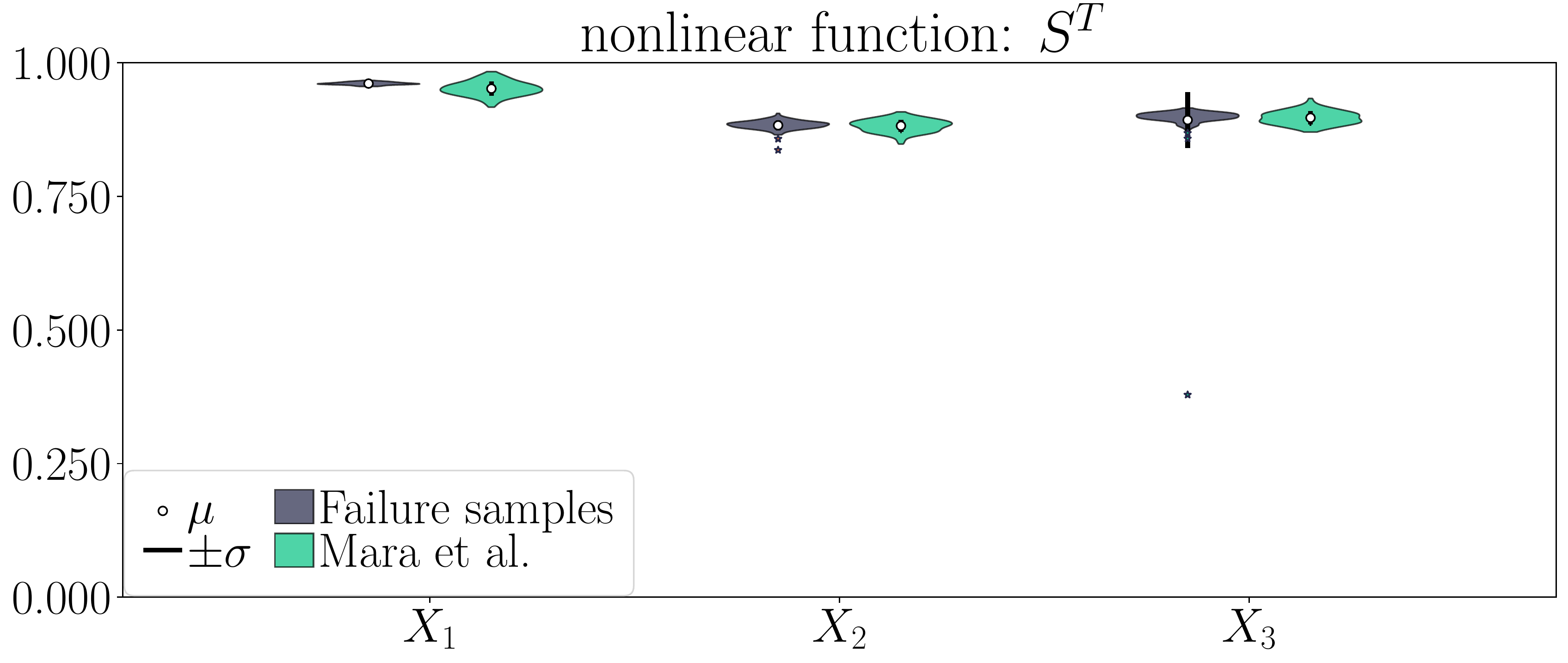}}
	\caption{Variance-based sensitivity indices for dependent inputs computed in 100 repeated runs with our approach (Failure samples; samples obtained with MC) and two reference methods (Mara et al. and Kucherenko et al.; both based on pick-freeze estimators) for the nonlinear test function.}
	\label{fig:nonlinfun_results}
\end{figure}

For our first example, we analyze a non-linear non-monotonic LSF that is considered in \cite{Li2019}:
\begin{align*}
	g(\bm{X}) &= X_1^3 + 10 X_2^2 + 0.1 \sin (\pi X_2) + 10 X_3^2 \\&+ 40 \sin (\pi X_3) +38,
\end{align*}
where the input is taken to be Gaussian:
\begin{equation*}
	\bm{X} \sim \mathcal{N}\left(\begin{bmatrix} 0 \\ 0 \\ 0\end{bmatrix}, \begin{bmatrix} 1 & 0.5 & 0.3 \\ 0.5 & 1 & 0.8 \\ 0.3 & 0.8 & 1\end{bmatrix}\right).
\end{equation*}
The reference failure probability is $\mathbb{P}(\mathrm{F}) = 5.34 \cdot 10^{-3}$ with 99 \% credible interval $[5.28,5.40] \cdot 10^{-3}$ based on $10^7$ independent samples.
We estimate the sensitivity indices with the proposed approach using failure samples generated with MC, the improved cross entropy method (iCE) \cite{Papaioannou2019b} and subset simulation (SUS) \cite{Au2001}.
\cref{fig:nonlinfun_results} compares estimator statistics (from 100 repeated runs) for all four variance-based RSA indices obtained with the failure samples-based approach (FS) using MC as well as two estimators based on applying the methods of \cite{Kucherenko2012} and \cite{Mara2016} to the indicator function of the failure domain. The FS estimators for independent Sobol' and Sobol' indices are unbiased and exhibit significantly lower variance than the benchmark solutions. The FS estimator for the total Sobol' indices of $X_2$ and $X_3$ produce some outliers, however the overall estimator variance is still reasonably confined (compare \cref{tab:luyis}). Moreover, the FS estimator of the independent total Sobol' index of $X_2$ exhibits a small bias. Generally, the FS estimators for total Sobol' indices or higher-order Sobol' indices are less accurate than their first-order counterparts due to the necessity of constructing multvariate KDEs of the failure-conditioned input PDFs. We suspect these results can be improved by allowing non-zero off-diagonal entries in the bandwidth matrix $\mathbf{H}$ while at the same time replacing the plug-in estimate used in \cref{eq:kde} with a more robust bandwidth selection method, e.g., based on cross-validation \cite{Duong2005}. 
\\~\\
\cref{tab:luyis} shows that the total computational cost of the FS estimator primarily depends on the RA approach that is used to generate failure samples. In particular, using advanced sampling methods such as the improved cross-entropy method (iCE) \cite{Papaioannou2019b} and SUS, the total number of LSF evaluations can be reduced by around two orders of magnitude compared to the MC-based pick-freeze approaches in \cite{Kucherenko2012,Mara2016} without compromising accuracy (compare estimator c.o.v. in \cref{tab:luyis}).  We remark that, while the number of failure samples was chosen equal to the number of samples per level for all examples in this section, it is generally possible to sample an arbitrary number of failure samples after completing the last level of either iCE or SUS.
\\~\\
The Sobol' indices of $X_2 $ and $X_3$ are larger compared to the respective independent Sobol' indices wheras the Sobol' index of $X_1$ is smaller than its independent counterpart. This is due to strong negative correlation between the $x_1$- and $x_2$-coordinates of the failure samples $\mathbf{X}$ (they have empirical correlation coefficient $-0.66$). Thus, the dependence structure of $X_1$ and $X_2$ effectively lowers $\mathbb{V}[\mathbb{P}(\mathrm{F})|U_1^{\{1\}}]$ compared to  $\mathbb{V}[\mathbb{P}(\mathrm{F})|U_1^{\{2\}}]$. While the variables ranking remains unchanged by the dependence structure, $X_1$ is signficantly more important than $X_2$ when considering only the computational model $\mathcal{Y}$ whereas both are almost equally important when considering the combination of both probabilistic and computational model.
\begin{table*}[]
	\setlength{\tabcolsep}{0.18cm}
	\centering
	\caption{Nonlinear test function: comparing computational cost and accuracy via total number of LSF calls and estimator c.o.v. as estimated over 100 repeated runs of each method using $10^6$ independent MC samples for Mara, Kucherenko and the MC-based failure-samples (FS) method and $10^4$ samples per level (a total of 4 levels are required in this example) for the iCE- and SUS-based FS methods. Estimates have been clipped to the interval $[0,1]$ to exclude nonsensical values (in this example, the only affected methods are the ones by Mara and Kucherenko).}
	\label{tab:luyis}
	\begin{tabular}{|l|c|c|c|c|c|c|c|c|c|c|c|c|c|}
		\hline
		method & effort &\multicolumn{12}{c|}{c.o.v.}\\
		& (total \# of &\multicolumn{3}{c|} {$S_{\mathrm{ind}}$} & \multicolumn{3}{c|} {$S$} & \multicolumn{3}{c|} {$S^T_{\mathrm{ind}}$} & \multicolumn{3}{c|} {$S^T$} \\  &LSF calls)&$X_1$&$X_2$&$X_3$ & $X_1$&$X_2$&$X_3$& $X_1$&$X_2$&$X_3$&$X_1$&$X_2$&$X_3$\\ 
		\hline
	 Mara & $7.5 \cdot 10^6$ & 0.16 & 0.59 & 0.74 & 0.23 & 0.70 & 0.26 & 0.01 & 0.02 & 0.01 & 0.01 & 0.01 & 0.01\\
	Kucherenko & $2.5 \cdot 10^6 $ &- &  - &  - & 0.09 & 0.38 & 0.11 & 0.02 & 0.02 & 0.02 & - &  - &  -\\
	FS (MC) & $1 \cdot 10^6$ &0.03 & 0.02 & 0.03 & 0.03 & 0.02 & 0.02 & 0.00 & 0.04 & 0.02 & 0.00 & 0.01 & 0.06\\
	FS (iCE) & $4 \cdot 10^4$ &0.03 & 0.02 & 0.02 & 0.03 & 0.02 & 0.02 & 0.01 & 0.05 & 0.02 & 0.00 & 0.03 & 0.01\\
	FS (SUS) & $4 \cdot 10^4$ &0.18 & 0.11 & 0.11 & 0.18 & 0.10 & 0.08 & 0.02 & 0.11 & 0.04 & 0.01 & 0.05 & 0.07\\
		

		\hline
	\end{tabular}
\end{table*}
\subsection{Short column}
\label{sec:shortcol}
In this section, we analyze a short column subjected to biaxial bending moments $M_1$ and $M_2$ and an axial force $P$ that was previously investigated in \cite{DerKiureghian2005}. The LSF reads
\begin{align*}
	g(\bm{X}) &= 1 - \frac{M_1}{s_1 Y} -\frac{M_2}{s_2 Y} - \left(\frac{P}{AY}\right)^\theta,
\end{align*}
where $\theta=2$ is a LSF parameter, $s_1 = 0.03\mathrm{m}^3$, $s_2 = 0.015\mathrm{m}^3$ are the flexural moduli of the short column section and $A = 0.190\mathrm{m}^2$ is the area of its cross section.
$Y$ is the yield strength of the column material and -- along with $M_1$, $M_2$ and $P$ -- is modelled as random variable. We use the Nataf model to define the joint distribution of these inputs following the specification of marginal distributions and the correlation structure in \cref{tab:shortcol}.
The reference failure probability is $\mathbb{P}(\mathrm{F}) = 9.29 \cdot 10^{-3}$ with 99 \% credible interval $[9.21,9.36] \cdot 10^{-3}$ based on $10^7$ independent samples. 
\begin{table}[]
	\setlength{\tabcolsep}{0.075cm}
	\centering
	\caption{Probabilistic model of short column example, including correlation matrix $\mathbf{R}_X$.}
	\label{tab:shortcol}
	\small{
		\begin{tabular}{llllcccc}
			\hline
			Input & Distr. & Mean & c.o.v. & \multicolumn{4}{c} {$\mathbf{R}_{X}$} \\ \cline{5-8}  &&&&$M_1$&$M_2$&$P$&$Y$\\ \hline
			$M_1$ [kNm]      & normal       &    $250$  &  $0.3$  &$1.0$ &$0.5$ & $0.3$ & $0.0$                  \\
			 $M_2$ [kNm]    & normal       &    $125$  & $0.3$  &$0.5$ &$1.0$ & $0.3$ & $0.0$                        \\ 
			 $P$ [kN]  & Gumbel& $2500$ & $0.2$ &$0.3$ &$0.3$ & $1.3$ & $0.0$           
			\\ 
			 $Y$ [N/$\text{mm}^2$]  & Weibull& $40$ & $0.1$ &$0.0$ &$0.0$ & $0.0$ & $1.0$           
			\\ \hline
			&              &      &                    &                                
		\end{tabular}
	}
\end{table}

\begin{table*}[]
	\setlength{\tabcolsep}{0.075cm}
	\centering
	\caption{Short column: comparing computational cost and accuracy via total number of LSF calls and estimator c.o.v. as estimated over 100 repeated runs of each method using $10^6$ independent MC samples for Mara, Kucherenko and the MC-based failure sample (FS) method and $5 \cdot 10^4$ samples per level (4 levels total) for the iCE- and SUS-based FS methods. Estimates have been clipped to the interval $[0,1]$ to exclude nonsensical values.}
	\label{tab:shortcol}
	\begin{tabular}{|l|c|c|c|c|c|c|c|c|c|c|c|c|c|c|c|c|c|}
		\hline
		method & effort &\multicolumn{16}{c|}{c.o.v.}\\
		& (total \# of &\multicolumn{4}{c|} {$S_{\mathrm{ind}}$} & \multicolumn{4}{c|} {$S$} & \multicolumn{4}{c|} {$S^T_{\mathrm{ind}}$} & \multicolumn{4}{c|} {$S^T$} \\  &LSF calls)&$M_1$&$M_2$&$P$ & $Y$ & $M_1$&$M_2$&$P$ & $Y$ & $M_1$&$M_2$&$P$ & $Y$ &$M_1$&$M_2$&$P$ & $Y$ \\ 
		\hline

Mara & $12 \cdot 10^6$ & 0.81 & 0.66 & 0.24 & 0.06 & 0.24 & 0.22 & 0.10 & 0.06 & 0.02 & 0.02 & 0.01 & 0.01 & 0.01 & 0.01 & 0.01 & 0.01\\
Kucherenko & $3 \cdot 10^6$ & - &  - &  - &  - & 0.09 & 0.09 & 0.06 & 0.04 & 0.02 & 0.02 & 0.01 & 0.01 &  - &  - &  - &  -\\
FS (MC) & $1 \cdot 10^6$ & 0.07 & 0.04 & 0.05 & 0.03 & 0.68 & 0.04 & 0.04 & 0.03 & 0.12 & 0.10 & 0.08 & 0.03 & 0.04 & 0.11 & 0.05 & 0.03\\
FS (iCE)   & $4 \cdot 10^4$ & 0.76 & 0.10 & 0.12 & 0.04 & 0.12 & 0.07 & 0.07 & 0.04 & 0.18 & 0.19 & 0.17 & 0.07 & 0.12 & 0.11 & 0.08 & 0.07\\
FS (SUS)   & $4 \cdot 10^4$ & 0.20 & 0.20 & 0.23 & 0.16 & 0.13 & 0.13 & 0.17 & 0.16 & 0.35 & 0.34 & 0.15 & 0.12 & 0.10 & 0.19 & 0.09 & 0.12\\
		\hline
\end{tabular}
\end{table*}
\begin{figure}[!ht]
	\centering
	\subfloat{\includegraphics[width=.5\textwidth]{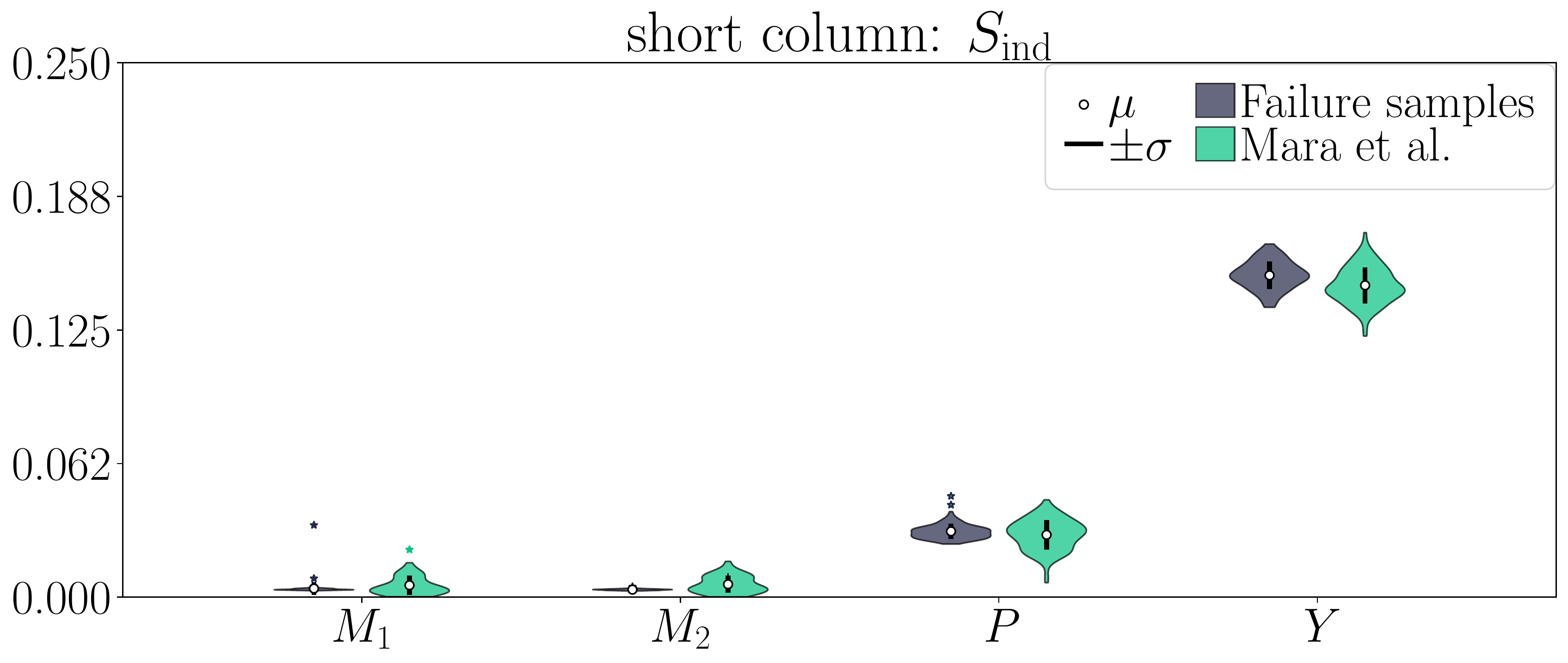}}\\
	\subfloat{\includegraphics[width=.5\textwidth]{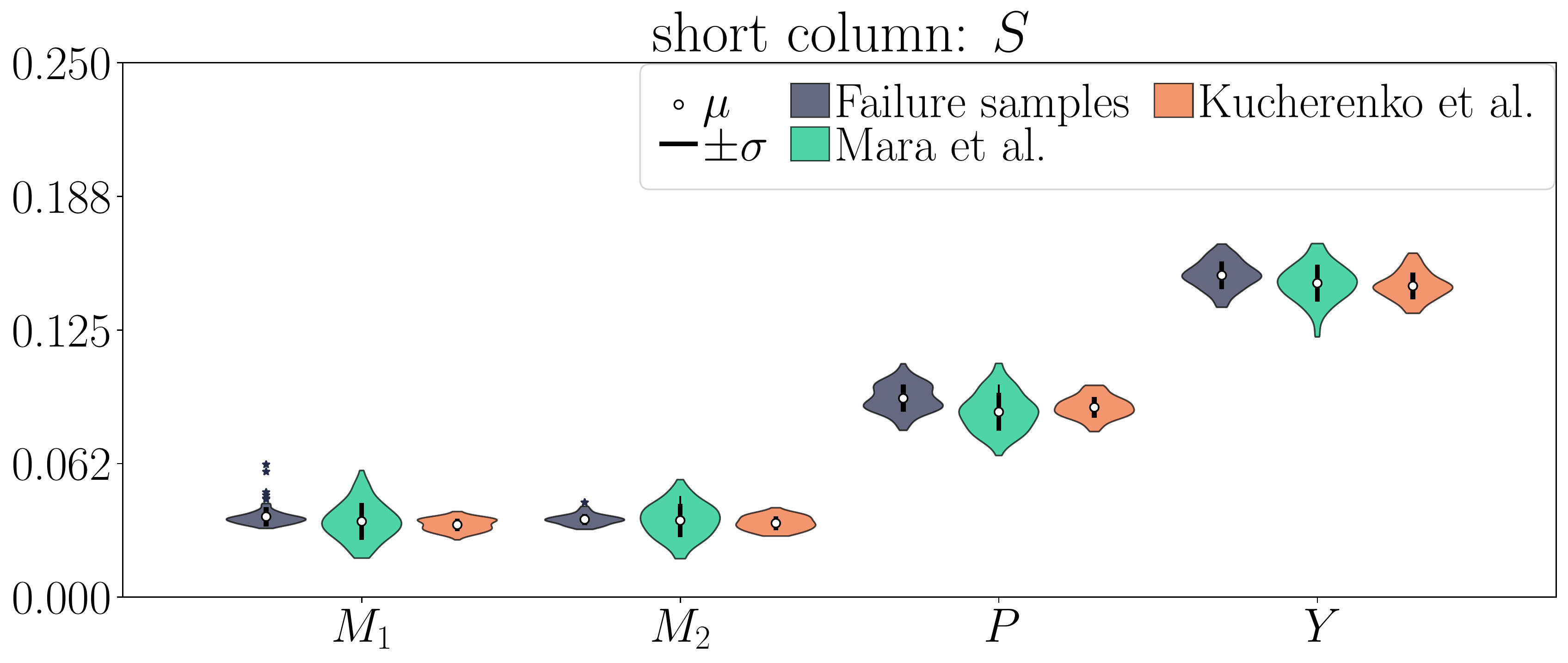}}\\
	\subfloat{\includegraphics[width=.5\textwidth]{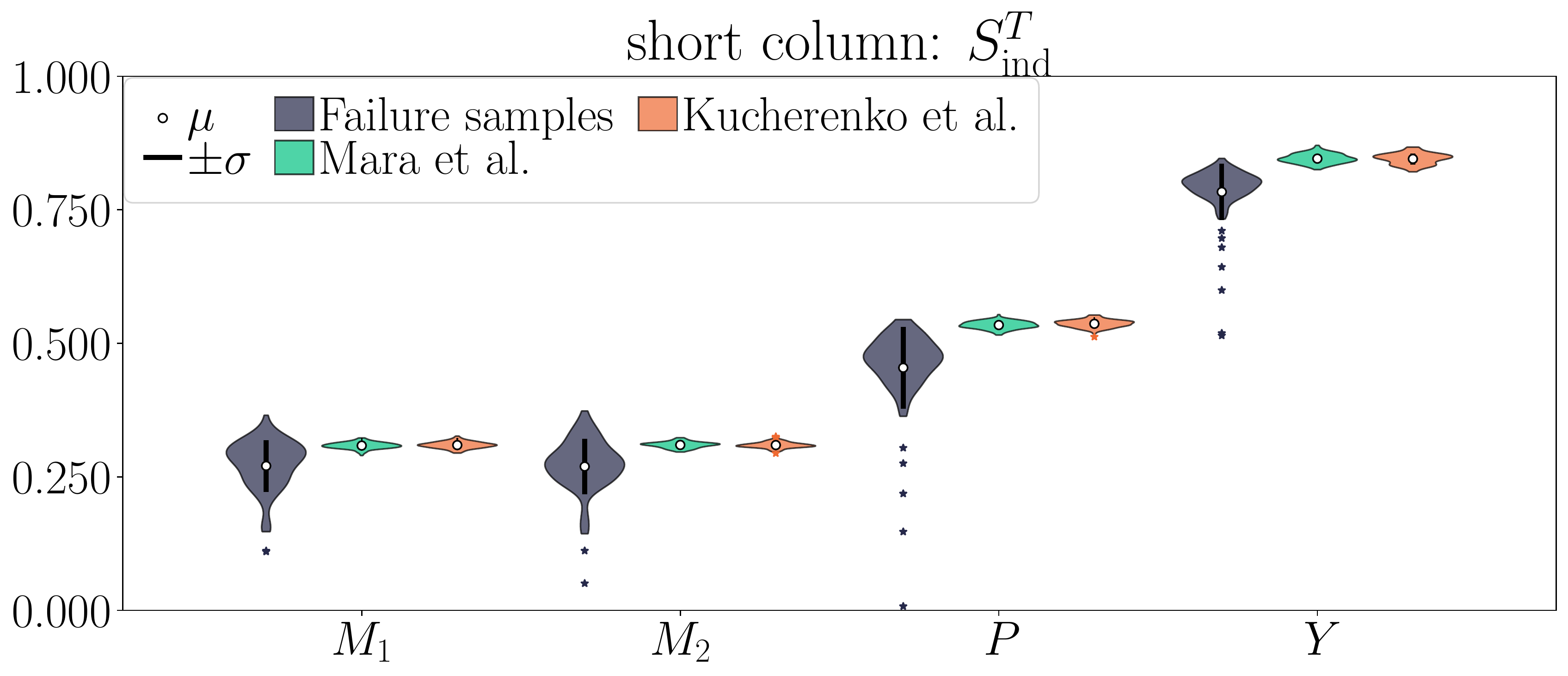}}\\
	\subfloat{\includegraphics[width=.5\textwidth]{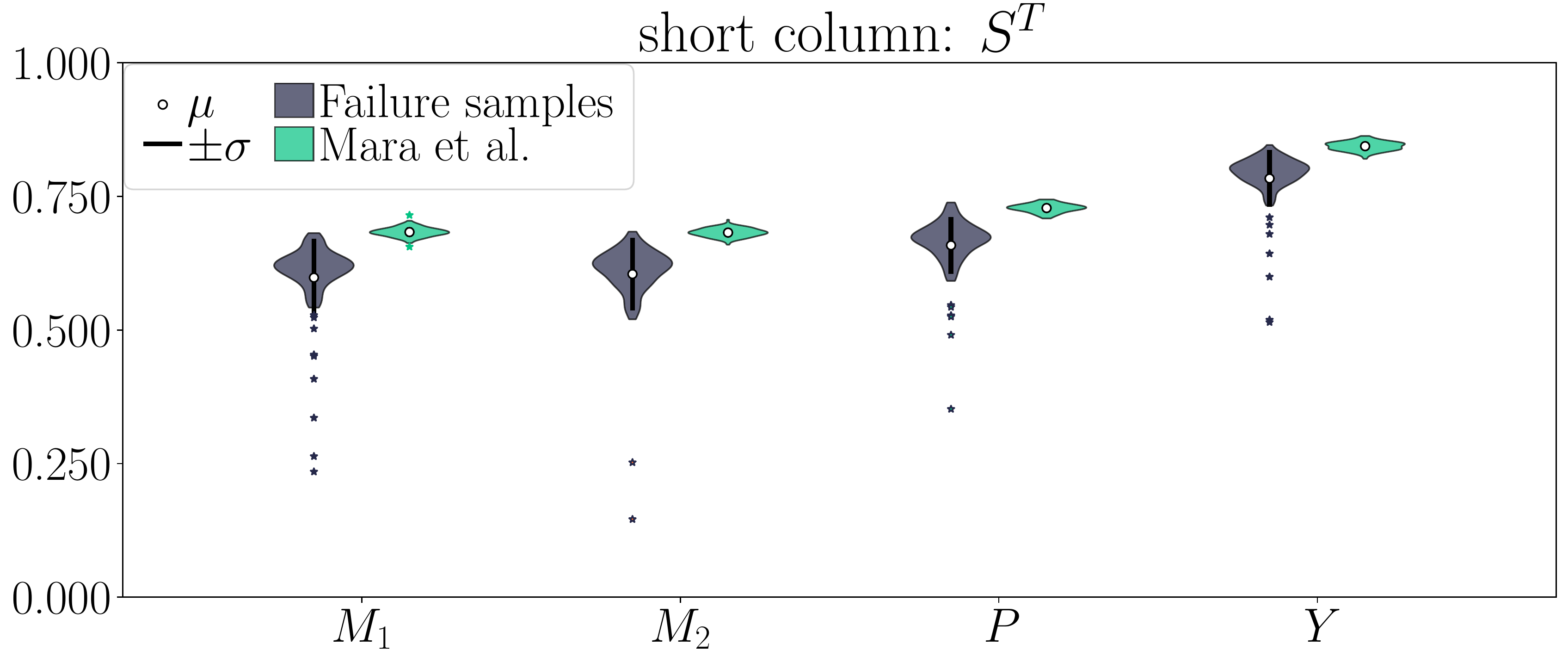}}
	\caption{Variance-based sensitivity indices for dependent inputs computed in 100 repeated runs with our approach (Failure samples; samples obtained with iCE) and two reference methods (Mara et al. and Kucherenko et al.; both based on pick-freeze estimators) for the short column example. }
	\label{fig:shortcol_results}
\end{figure}
~\\
\cref{fig:shortcol_results} shows a comparison of the full set of indices computed with both reference methods and the FS approach using iCE. The rankings produced by any of the indices coincide: the yield strength $Y$ is the most important variable followed by the load $P$ and the bending moments $M_1$ and $M_2$, which are equally unimportant.  The FS-iCE estimator variability for independent Sobol' and Sobol' indices is on par with or smaller than that of the reference methods. For the total Sobol' indices, the FS-iCE estimator produces larger variability with some notable outliers. This performance decrease is also reflected in the estimator c.o.v. documented in \cref{tab:shortcol} and underpins the fact that the FS estimator is not well-adapted to compute higher-order or total indices outside low-dimensional models (here, four input dimensions are sufficient for significant inaccuracies to arise in the associated three-dimensional KDEs based on the available number of failure samples). As mentioned above, refining the KDE approach, which is the major source of error in the computation of these total Sobol' index estimates, may alleviate these problems but as soon as $d$ becomes truly large, the FS estimator is not up to the task. 
\subsection{Monopile foundation}
\label{sec:monopile}
\begin{figure}[!ht]
	\centering
	\includegraphics[width=.425\textwidth]{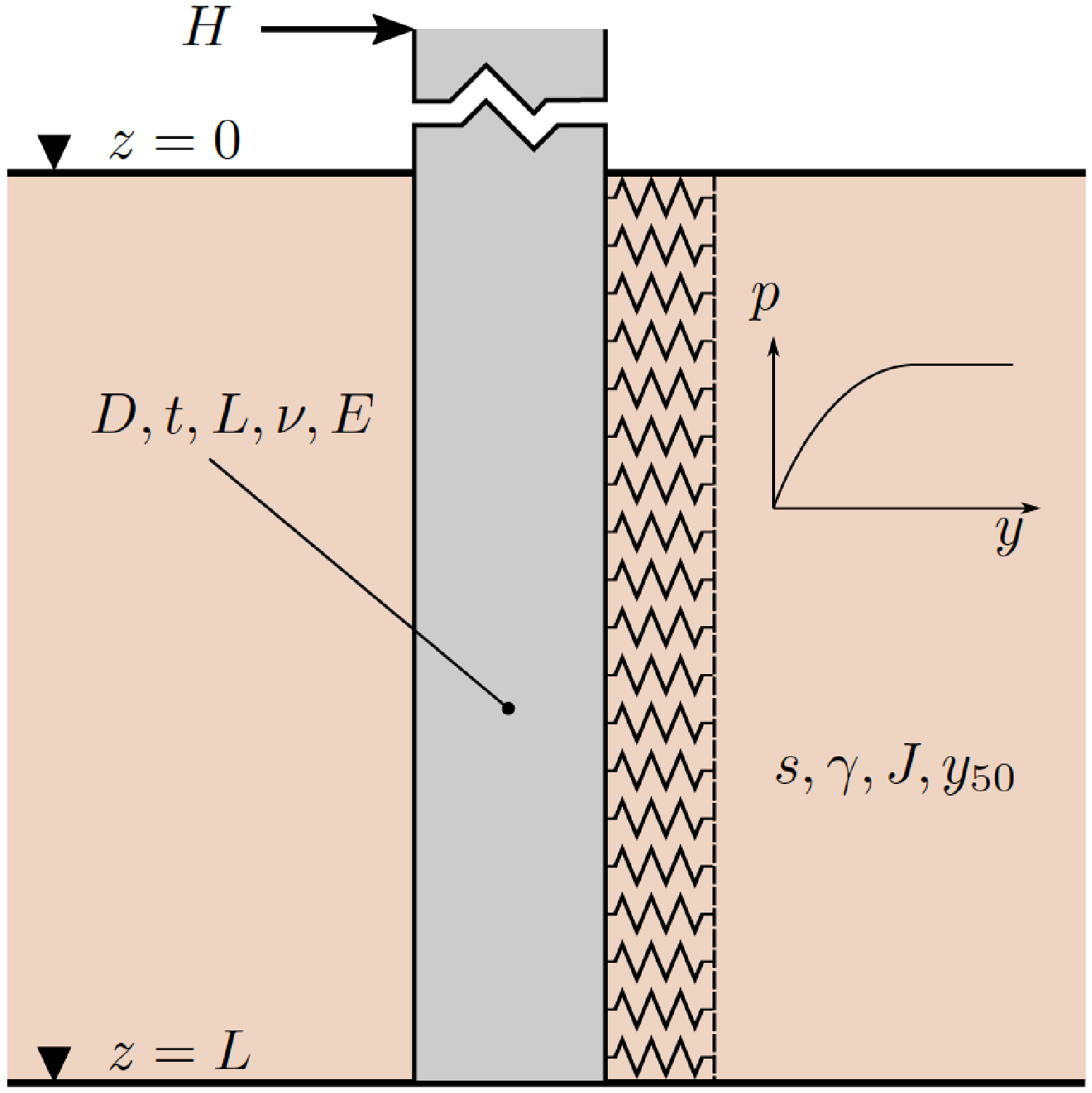}
	\caption{Wind turbine monopile foundation.}
	\label{fig:monopile}
\end{figure}

We consider the finite element model of the concrete monopile foundation of an offshore wind turbine (Figure \ref{fig:monopile}) as it interacts with stiff, plastic soil. The monopile has a depth of $L = 30$ m, a diameter of $D = 6$ m, a wall thickness of $t=0.07$ m. Deterministic parameters of the concrete are its Poisson ratio $\nu = 0.3$ and its Young's modulus $E = 2.1\cdot10^{5}$ MPa. Uncertain parameters comprise the lateral load $H$ and the undrained shear strength of the soil $s$ as well as hyperparameters of both quantities. The engineering model setup follows \cite{Depina2017} and the probabilistic model considered there has been modified following \cite{Jiang2018}.
$s$ is modelled by a random field with linear mean drift along the soil depth coordinate $z$. Given an underlying stationary Gaussian random field $\tilde{s}(z,\bm{X})$
\begin{equation}
	\tilde{s}(z) \sim \mathcal{N}(0,\sigma_{\tilde{s}}) ~\text{for}~0 \le z \le L,
\end{equation}
the non-stationary random field representing the shear strength of the soil can be expressed as
\begin{equation}
	s(z,\bm{X}) = s_0 + s_1 \gamma z \exp\left\{ \tilde{s}(z,\bm{X})\right\},
\end{equation}
where $\gamma = 18 \mathrm{kN/m^3}$ is the soil unit weight, $s_0$ is the undrained shear strength at ground level and $s_1$ is the drift parameter governing the mean increase of $s$ with increasing soil depth.
$\tilde{s}(z,\bm{X})$ models the intra-site variability; at a given site with known $s_0$ and $s_1$, it describes the inherent variability of $s$. In order to describe the inter-site variability of $s$, the parameters $s_0$ and $s_1$ are uncertain as well (see \cref{tab:monopile_inputs}).
The stationary RF $\tilde{s}$ is modelled using an exponential-type correlation function
\smash{$\rho_{\tilde{s}\tilde{s}}(z',z'') = \exp\left\{ -(2 |z'-z''|)/\theta_{\tilde{s}} \right\}$}
with vertical soil scale of fluctuation $\theta_{\tilde{s}} = 1.9$m and $\sigma_{\tilde{s}} = 0.3$.
$\tilde{s}$ is discretized through application of the midpoint method \cite{DerKiureghian1988} in terms of an 82-dimensional standard Gaussian random vector $\bm{\xi}$.
The load $H$ acting on the foundation is modelled with a Gumbel distribution with log-normally disitributed location and scale parameters $A$ and $B$. 
The full probabilistic input model is summarized in \cref{tab:monopile_inputs}.
The output of the finite element model is the stress field of the foundation and failure is assumed to occur as if the maximally occurring stress exceeds $80 MPa$.
A more detailed account of the probabilistic modelling step and the choice of several deterministic parameters above is given in \cite{Ehre2020}.
\begin{table}[!ht]
	\caption{ Input variable definitions of the monopile.}
	\label{tab:monopile_inputs}
	\centering
	\begin{tabular}{p{.15cm}lccc}
		\hline
		Input & & Dist. & Mean $\mu$  & c.o.v. $\delta$\\
		\hline
		$\bm{\xi}$&[-]& Std-Normal & $\bm{0}$ & - ($\bm{\Sigma} = \bm{I}$)\\
		$s_0$ &[kPa]& Log-Normal & $33.7094$ & $0.3692$\\
		$s_1$ &[kPa]& Log-Normal & $0.7274$ & $0.8019$\\
		$H$&[kN]& Gumbel & \smash{$\mu_{H|A,B}$} & \smash{$\delta_{H|a_H,b_H}$}\\
		$A$& [kN] & Log-Normal & $2274.97$ & $0.2$ \\
		$B$&[kN]& Log-Normal &  $225.02$&  $0.2$\\
		\hline
	\end{tabular}
\end{table}
\\~\\
We compute the independent Sobol' and Sobol' indices of the parameters $s_0, s_1,A,B$ and $H$. Analyses of the monopile model in \cite{Ehre2020,Straub2022} have shown the discretized underlying Gaussian random field $\bm{\xi}$ to have negligible effect on system failure such that we don not evaluate the sensitivity  with respect to $\bm{xi}$\footnote{While non-sensical, it would also be technically impossible to compute the  82nd-order Sobol' index representing the effect of $\bm{\xi}$ as this requires $82$-dimensional KDEs in the FS approach. This will not yield reliable sensitivity estimates based on the results for the short column example, where three-dimensional KDEs already produced significant inaccuracies in the sensitivity estimates}. Due to the independence of $s_1$ and $s_0$ we have $S = S_{\mathrm{ind}}$ for both of these variables. The load variables $H,A,B$ are modelled with the Rosenblatt transform and a generic joint distribution. The latter is given in the form of conditional disitributions that relate to the ordering $\bm{X} = [A,B,H]$ (see \cref{tab:monopile_inputs}). Hence, obtaining \smash{$T^{\{1\}}=\Phi^{-1} \circ [F_{A},F_{B},F_{H|A,B}]$} is straight-forward. However, in order to obtain  \smash{$T^{\{2\}}=\Phi^{-1} \circ [F_{B},F_{H|B},F_{A|H,B}]$} for the ordering $\bm{X} = [B,H,A]$ and \smash{$T^{\{3\}}=\Phi^{-1} \circ [F_{H},F_{A|H},F_{B|A,H}]$} for the ordering $\bm{X} = [H,A,B]$, we require several additional marginal and conditional CDFs. First we simplify the expressions for the required CDFs as far as possible in terms of $f_A$, $f_B$ and $f_{H|A,B}$ since these have standard distribution types and are hence convenient to evaluate. With PDF and CDF subscripts dropped in order to keep the resulting formulae uncluttered, these read
\begin{align}
	 F(h|b) &=  \int_0^\infty F(h|a,b) f(a) \mathrm{d}a \\
	 F(a|h,b) &= \frac{\int_{0}^a f (h|a',b) f(a') \mathrm{d}a'}{\int_{0}^\infty f (h|a',b) f(a')   \mathrm{d}a'}\\
	 F(h) &=  \int_{0}^\infty \int_0^\infty F (h|a,b) f(a) f(b)  \mathrm{d}a \mathrm{d}b \\		
	 F(a|h) &=\frac{\int_{0}^a \int_{0}^\infty f (h|a',b) f(a')  f(b) \mathrm{d}b\mathrm{d}a'}{\int_{0}^\infty \int_{0}^\infty f (h|a',b) f(a')  f(b) \mathrm{d}b\mathrm{d}a'}\\
	 F(b|a,h) &= \frac{\int_{0}^b  f (h|a,b') f(b') \mathrm{d}b'}{\int_{0}^\infty f (h|a,b')  f(b') \mathrm{d}b'}.
 \end{align}
The above expressions are readily approximated using numerical or MC integration schemes at any triplet $(h,a,b)$. It is clear that using the Rosenblatt transformation for this purpose can become quite tedious if the number of variables increases beyond $d=3$. Usually, exactly one transformation arises naturally from the modelling step and then $d -1$ others have to be determined in the above manner, i.e., by solving for the associated $d(d-1)$ conditonal CDFs.
Not only can this become tedious but as $d$ increases the integrals involved in this step may become intractable. Hence using the approach of \cite{Mara2016} without imposing a Nataf model is only reasonable for small sets of dependent variables such as $A,B,H$ in this example.
\begin{figure}[!ht]
	\centering
	\subfloat{\includegraphics[width=.5\textwidth]{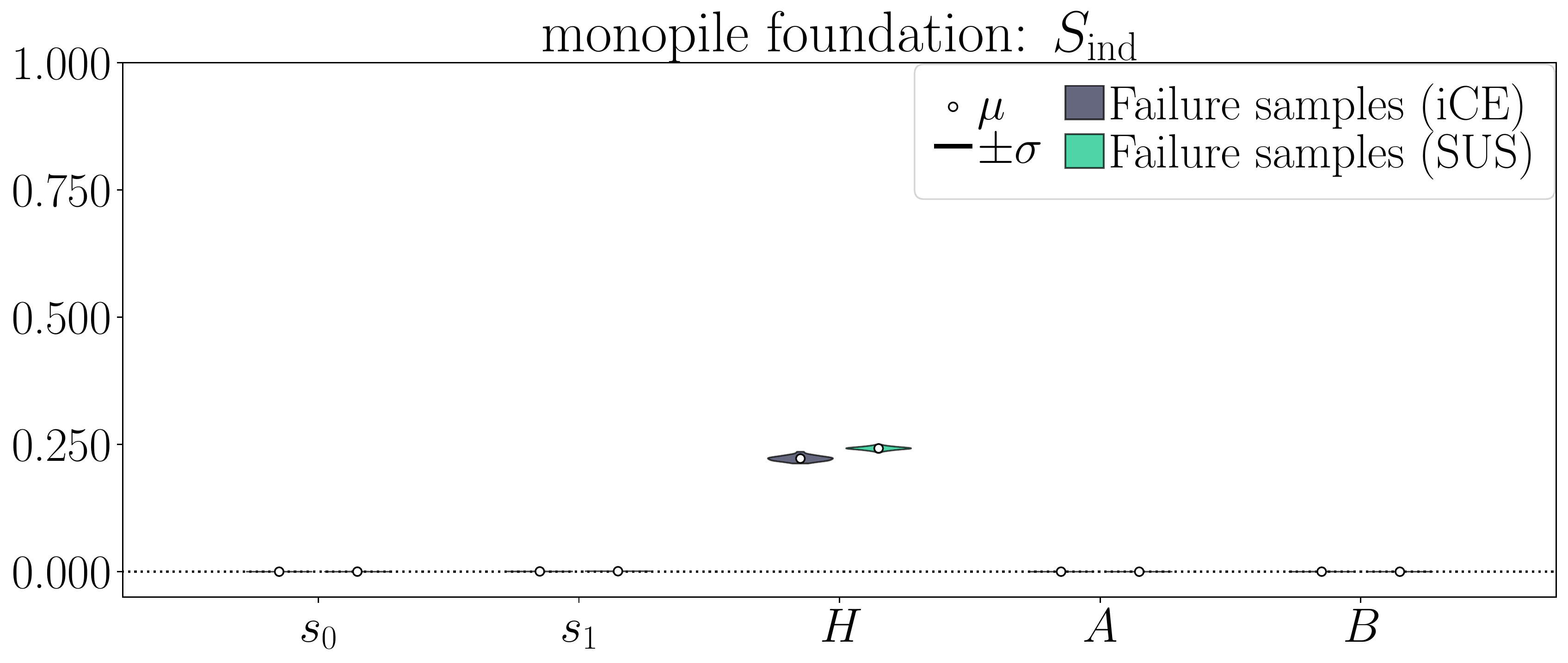}}\\
	\subfloat{\includegraphics[width=.5\textwidth]{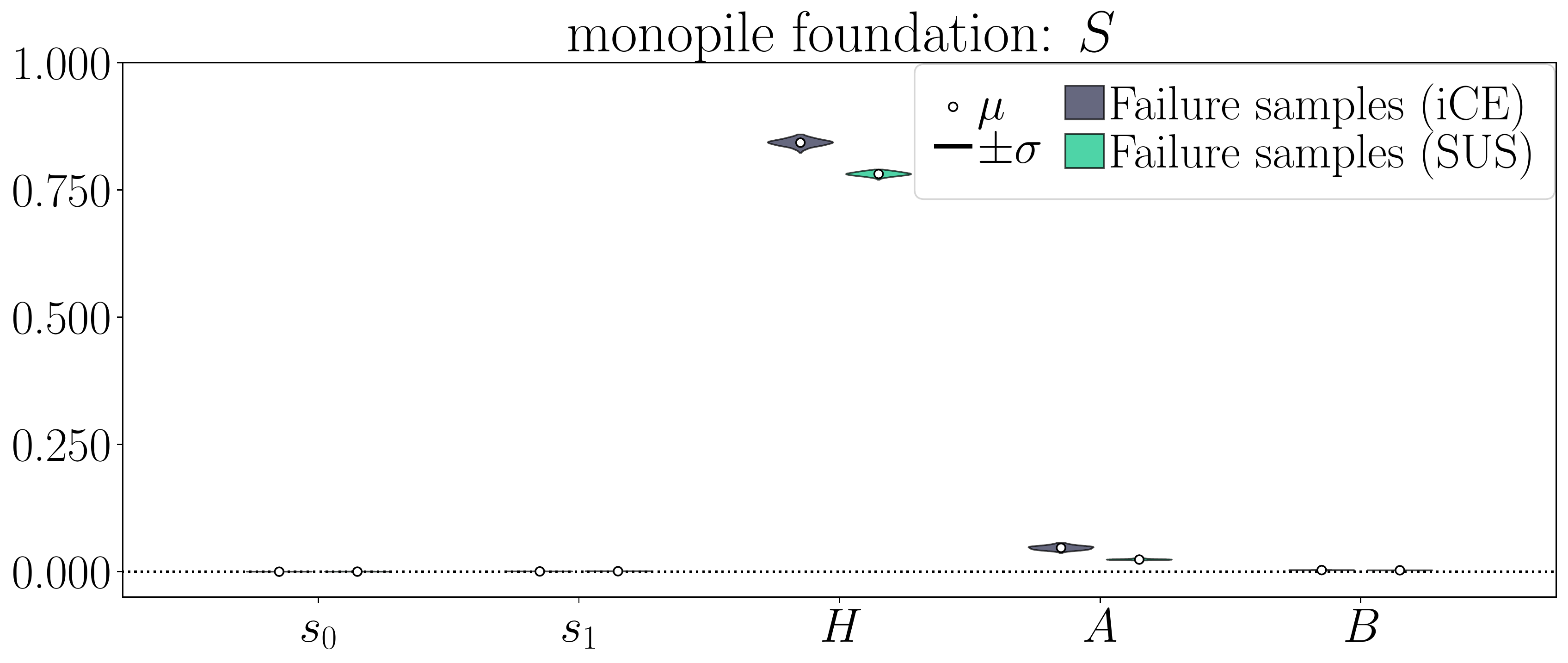}}
	\caption{Variance-based sensitivity indices for dependent inputs computed based on bootstrapping ($100$ copies) a single set of $10^4$ failure samples (one per method). Due to independence, $S = S_{\mathrm{ind}}$ for $s_0,s_1$.}
	\label{fig:monopile_results}
\end{figure}
\\~\\
We estimated independent Sobol' and Sobol' indices with FS-iCE and FS-SUS using $10^4$ and $3 \cdot 10^4$ samples per level. With each method we generated a single set of of $10^4$ and $3 \cdot 10^4$ failure samples, respectively. Estimator statistics in this case are formed with a 100-fold bootstrap of sample size $10^4$ in both methods and are illustrated in \cref{fig:monopile_results} and \cref{tab:monopile}.\footnote{We bootstrap over a larger number of failure samples with SUS since -- as opposed to iCE -- SUS produces dependent failure samples. }
\\~\\
\cref{fig:monopile_results} shows full and independent first-order Sobol' indices for $H$, $A$, $B$, $s_0$ and $s_1$ computed with both FS-SUS and FS-iCE. Both methods produce similar sensitivity estimates for each of the inputs. 
In \cref{tab:monopile}, we report the bootstrap-based  empirical estimator c.o.v. of these two methods and find they are of comparable accuracy.
The full first-order Sobol' indices reported in \cref{fig:monopile_results} (bottom) indicate that the indicator function is relatively sensitivite to changes in wind load-related inputs whereas soil material-related inputs bear  negligible effects regarding failure. This is consistent with the analysis of the monopile model given in \cite{Ehre2020}. The independent first-order Sobol' indices of the load inputs  \cref{fig:monopile_results} (top) reveal that the wind load $H$ still has significant influence on the failure event through the LSF only at $S_{\text{ind},H} \approx 0.25$. However, comparing this value to the full index of $H$, $S_{H} \approx 0.75$, approximately two thirds of the load influence are contributed through the probabilistic model by virtue of the dependence of $H$ on $A$ and $B$. 
The full indices of $A$ and $B$ are small compared to that of $H$ but not zero. In \cite{Papaioannou2021b}, the behaviour of first-order and total Sobol' indices of LSF indicator functions at different failure probability magnitudes are studied: first-order effects decrease with decreasing failure probability magnitude such that a larger portion of the indicator variance is explained by higher-order effects. 
The independent first-order Sobol' indices of $A$ and $B$ are 0. This is to be expected as they only influence the model output (and the failure event therefore) through the probabilistic model and do not enter the LSF directly.
\begin{table*}[]
	\setlength{\tabcolsep}{0.075cm}
	\centering
	\caption{Monopile: comparing computational cost and accuracy via total number of LSF calls and bootstrapped estimator c.o.v. using 100 resampled failure sample sets and $10^4$ samples per level (4 levels total) for the iCE- and SUS-based FS methods.}
	\label{tab:monopile}
	\begin{tabular}{|l|c|c|c|c|c|c|c|c|c|c|c|}
		\hline
		method & effort &\multicolumn{10}{c|}{c.o.v.}\\
		& (total \# of &\multicolumn{5}{c|} {$S_{\mathrm{ind}}$} & \multicolumn{5}{c|} {$S$}\\  &LSF calls)&$s_0$&$s_1$&$H$ & $A$ & $B$&$s_0$&$s_1$&$H$ & $A$ & $B$ \\ 
		\hline

			FS (iCE)   & $4 \cdot 10^4$  & 0.14 & 0.07 & 0.04 & 0.04 & 0.51 & 0.14 & 0.08 & 0.01 & 0.02 & 0.02\\
			FS (SUS) & $12 \cdot 10^4$ & 0.21 & 0.07 & 0.02 & 0.02 & 0.19 & 0.21 & 0.06 & 0.01 & 0.01 & 0.01\\
		\hline
	\end{tabular}
\end{table*}
\FloatBarrier
\section{Concluding remarks}
\label{sec:concluding_remarks}
In this contribution, we propose to analyze the effect of input parameter dependence in the framework of variance-based reliability sensitivity analysis. \cite{Mara2012,Mara2016} first suggested computing an extended set of Sobol' and total Sobol' indices that distinguishes between variables affecting the output through the computational model directly and variables affecting the output indirectly by affecting each other through the probabilistic model (dependence). This approach is suitable for generic model output. In this contribution, we extend the approach to the reliability sensitivity setting by computing an extended set of Sobol' indices for a rare event indicator function target. We devise an estimation procedure that relies on a set of failure samples only. As such, the method comes at almost no additional computational cost once a set of failure samples has been obtained, e.g., as the byproduct of a sample-based rare event estimation method. This is in contrast to the originally proposed way of computing these indices, where $d$ repeated sensitivity analyses and sampling steps have to be performed where $d$ is the number of input parameters. 
\\~\\
While SA methods that are not focussed on rare events usually deteriorate in the RSA setting, it is easily possible to extend our RSA approach back to a more general SA setting where the sensitivity target is not related to a rare event but merely the model output $Y$: replacing failure samples with $Y$-samples, an extended set of Sobol' indices for $Y$ may be obtained by using a binning method to estimate the underlying empirical conditional expectations \cite{LiMahadevan2016}.  It is thereby possible to reduce the computational cost by a factor $d$ in the context of SAMO as well. However, in this case -- as well as for the reliability sensitivity approach presented in this paper -- the estimators will suffer from the curse of dimensionality meaning that they are well suited for computing first-/low-order Sobol' indices irrespective of $d$ and total Sobol' indices if $d$ is low only.
\\~\\
We demonstrate the efficacy of the proposed aproach in computing first-order Sobol' indices on a strongly nonlinear test function as well as a low- and a high-dimensional engineering example. With our approach, we are able to reduce the required number of model evaluations by over two orders of magnitude compared to using pick-freeze-based methods such as \cite{Mara2016} and \cite{Kucherenko2012}. We use the Rosenblatt transform both with generic distribution models and the Nataf model to formulate the probabilistic model and point out the drawback associated with using generic models for computing the full set of indices.

\section*{Acknowledgments}
We acknowledge support by the German Research Foundation (DFG) through Grant PA 2901/1-1.






\bibliographystyle{elsarticle-num}
\bibliography{mybib}







\end{document}